\documentclass[11pt]{article}
\usepackage[margin=1in]{geometry}
\usepackage{amsmath}
\usepackage{amssymb}
\usepackage{amsthm}
\usepackage{mathtools}
\usepackage{xcolor}
\usepackage{microtype}
\usepackage[backend=biber,isbn=false,date=year,maxnames=99]{biblatex}
\usepackage[colorlinks=true]{hyperref}
\hypersetup{
    linkcolor = {violet},
    citecolor = {violet},
    urlcolor = {teal}
}

\newenvironment{acks}{\section*{Acknowledgments}}{}
\newtheorem{theorem}{Theorem}[section]
\newtheorem{conjecture}[theorem]{Conjecture}

\newcommand\Set[1]{\{#1\}}

\DeclareMathOperator{\Root}{root}
\newcommand{\Null}{\mathtt{null}}
\newcommand{\Level}{\ell}
\newcommand{\Depth}{d}
\DeclareMathOperator{\Postorder}{postorder}

\newcommand{\Algorithm}{\mathcal{A}}
\newcommand{\Instance}{\mathcal{I}}
\newcommand{\Model}{\mathcal{M}}

\newcommand{\Simulate}{\mathcal{S}}
\newcommand{\Repeat}{\mathcal{F}}
\newcommand{\Reduce}{\mathcal{R}}
\newcommand{\Graph}{\mathcal{G}}

\DeclareMathOperator{\Cost}{cost}
\DeclareMathOperator{\OPT}{OPT}
\newcommand{\OPTmut}{\OPT_{\operatorname{mut}}}
\newcommand{\OPTMin}{\OPT_{\min}}
\newcommand{\OPTrot}{\OPT_{\operatorname{rot}}}
\newcommand{\CrossingBound}{\Lambda}
\newcommand{\Wilber}{\Lambda_2}

\DeclareMathOperator{\Splay}{splay}
\DeclareMathOperator{\TopDownSplay}{top-down-splay}
\DeclareMathOperator{\MoveToRoot}{move-to-root}

\newcommand{\Insert}{\mathtt{insert}}
\newcommand{\Delete}{\mathtt{delete}}
\newcommand{\Search}{\mathtt{search}}

\newcommand{\Push}{\mathtt{push}}
\newcommand{\Pop}{\mathtt{pop}}
\newcommand{\Inject}{\mathtt{inject}}
\newcommand{\Eject}{\mathtt{eject}}

\DeclareMathOperator{\BST}{BST}

\title{A Foundation for Proving Splay is Dynamically Optimal\footnote{This paper is an adaptation of the first author's Ph.D.\ thesis~\cite{LEVY_THESIS}. We presented an earlier version at SODA~\cite{A_NEW_PATH}.}}

\author{
  Caleb C. Levy\footnote{Sunshine; \mbox{caleb.levy@gmail.com}.}\\
  \and
  Robert E. Tarjan\footnote{Department of Computer Science, Princeton University; Intertrust Technologies; \mbox{ret@cs.princeton.edu}.}
}

\date{}

\begin{document}
\maketitle
\begin{abstract}
Consider the task of performing a sequence of searches in a binary search tree.
After each search, we allow an algorithm to arbitrarily restructure the tree.
The cost of executing the task is the sum of the time spent searching and the
time spent optimizing the searches with restructuring operations. Sleator and
Tarjan introduced this notion in 1985, along with an algorithm and a
conjecture. The algorithm, Splay, is an elegant procedure for performing
adjustments that move searched items to the top of the tree. The conjecture,
called dynamic optimality, is that the cost of splaying is always within a
constant factor of the optimal algorithm for performing searches. We lay a
foundation for proving the dynamic optimality conjecture. Central to our method
is approximate monotonicity. Approximately monotone algorithms are those whose
cost does not increase by more than a fixed multiple after removing searches
from the sequence. As we shall see, Splay is dynamically optimal if and only if
it is approximately monotone. This result extends to a weaker form of
approximate monotonicity as well as insertion, deletion, and related
algorithms. We prove that a lower bound on optimal execution cost is
approximately monotone and outline how to adapt this proof from the lower bound
to Splay, and how to overcome the remaining barriers to establishing dynamic
optimality.
\end{abstract}

\section{Context}\label{sec:Context}

The binary search tree is the canonical pointer-based data structure for
maintaining a sorted collection in fast memory. Its most attractive feature is
that the number of comparisons required to verify the presence of an item is
logarithmic in the size of the tree, provided that the tree is properly
arranged. Without exercising care when adding elements, however, a binary
search tree can easily become unbalanced, making search cost proportional to
the size of the tree in the worst case. Thus binary search trees require some
form of maintenance and restructuring for good performance.

Adel'son-Vel'skii and Landis gave the first method that guarantees efficient
searches in the presence of updates~\cite{AVL_TREES}. They supplement nodes
with bits that provide rough information about how balanced each node's
subtrees are. After an insertion or deletion, a restructuring procedure
restores invariants on the balance bits. These invariants ensure that all paths
in the tree have length at most logarithmic in the tree's size. There are many
variations of this idea. Perhaps most famous is the red-black tree due to its
requiring fewer restructuring operations and a catchy
name~\cite{RED_BLACK_TREES}. Restructuring schemes based on balance bits remain
an active topic of research~\cite{WEAK_AVL_TREES}. Many more schemes now exist.
Randomized search trees, such as treaps~\cite{TREAPS}, zip
trees~\cite{ZIP_TREES_JOURNAL} and others~\cite{RANDOM_BALANCE} trade
worst-case performance guarantees for good expected behavior in order to gain
simpler rebalancing procedures and fewer pointer changes. Scapegoat
trees~\cite{ANDERSSON_SCAPEGOAT_JOURNAL,RIVEST_SCAPEGOAT} defer restructuring
operations until they can be executed in bulk. B-trees~\cite{B_TREES} and their
derivatives~\cite{B_PLUS}, close relatives of binary search trees that use
system memory characteristics to determine node arity, are ubiquitous in
database applications. There are numerous related data structures. For the most
part, they are well-understood. However, there is a class of binary search tree
algorithm whose behavior remains one of the great open questions in theoretical
computer science.

While the above-mentioned data structures guarantee logarithmic search time,
they usually cannot perform much better than this. Real-world access patterns
often have some latent structure. For example records may be arranged in
partially sorted sub-blocks, and databases often receive frequent requests for
a small number of high-traffic elements. In such situations it can be possible
to do better than logarithmic time per access by adjusting the tree after
searches, instead of solely after adding or removing elements. This leads to
colloquially named ``self-adjusting'' binary search tree algorithms. Allen and
Munro were the first to examine such algorithms in depth~\cite{MOVE_TO_ROOT}.
They developed a simple procedure with good expected behavior in many cases. By
far the most famous self-adjusting binary search tree algorithm is Sleator and
Tarjan's improvement to this procedure, called Splay~\cite{SPLAY_JOURNAL},
which has many compelling properties and applications. For our purposes,
Sleator and Tarjan's most important contribution to this topic is not what they
proved, but instead what they left unresolved. The dynamic optimality
conjecture asserts that Splay is essentially the ideal algorithm for every
possible access pattern. This problem's intrigue arises from several sources.

Dynamic optimality would imply we can use Splay as a stand-in for many
more-specialized data structures. Splay simultaneously acts as a balanced
search tree and as a spatial and a temporal cache. It also shares properties
with entropy-minimizing static trees~\cite{SPLAY_JOURNAL} and data structures
for disjoint set union~\cite{DISJOINT_SETS}. Dynamically optimal algorithms can
emulate multi-finger binary search trees~\cite{MULTI_FINGER_TREES} and
doubly-ended queues~\cite{SEQUENTIAL_ACCESS}.

Advances in our understanding of binary search trees percolate into other
areas. Splay and related self-adjusting data structures inspired the creation
of pairing heaps~\cite{PAIRING_HEAPS}, smooth heaps~\cite{SMOOTH_HEAPS_JOURNAL}
and slim heaps~\cite{SMOOTH_HEAPS_ANALYSIS}. Splay's properties find utility in
encoding schemes~\cite{SPLAY_ENCODING}, routing problems~\cite{SPLAYNET} and
optimizing for concurrent non-uniform access~\cite{CB_TREE}, and the conjecture
has analogues for B-trees~\cite{B_TREE_DYNAMIC_OPTIMALITY,BELGA_B_TREES},
search-tree-on-tree data
structures~\cite{SPLAY_TREES_ON_TREES,STT_DYNAMIC_OPTIMALITY} and external
memory settings~\cite{EXTERNAL_MEMORY}.

Furthermore, the conjecture has become a nexus for the development of new
strategies for analyzing data structures. Splay was intimately involved in the
adaptation of potential functions from physics to computer
science~\cite{AMORTIZED_COMPLEXITY}. Concepts common in the analysis of
forbidden substructures are frequently applied to Splay and related
self-adjusting algorithms, with examples including Davenport-Schinzel
sequences~\cite{DAVENPORT_SCHINZEL}, forbidden
submatrices~\cite{FORBIDDEN_SUBMATRIX} and pattern-avoiding
permutations~\cite{PATTERN_AVOIDANCE,SIMPLER_PATTERN_AVOIDANCE}. Techniques
from computational geometry are now common in this area of
research~\cite{GEOMETRY,GREEDY_DYNAMIC_FINGER}.

Finally, the conjecture has a distinct intellectual allure. At first blush, its
claim seems too good to be true, which makes the idea of proving it all the
more attractive. Its statement is elegant and deceptively simple, yet anyone
who has attempted to tackle the problem can attest to its subtlety and utter
defiance of standard mathematical approaches. Solutions frequently seem
tantalizingly close while remaining just out of reach.

This investigation adopts a somewhat different tone from its companions. It can
often be easier to induct on stronger hypotheses because they provide more
exploitable structure. Accordingly, we have no qualms about presuming that
Splay is dynamically optimal and allowing this to guide our intuition. Our
objective is to determine \emph{how} we can prove the conjecture, not
\emph{if}. Section~\ref{sec:Preliminaries} defines our execution model and
summarizes related work. Section~\ref{sec:ApproximateMonotonicity} shows that
Splay is dynamically optimal if and only if it is approximately monotone.
Section~\ref{sec:StartupOverhead} formalizes optimality with additive overhead
and demonstrates that if Splay is optimal then it has no such overhead.
Section~\ref{sec:Mutation} extends both Splay and our execution model to
incorporate mutation operations and establishes that if Splay is optimal
without these operations then it is optimal when they are permitted.
Section~\ref{sec:NaturalAlgorithms} generalizes our results to similar
algorithms. Section~\ref{sec:CrossingCost} establishes that a non-trivial lower
bound on optimal execution cost is approximately monotone.
Section~\ref{sec:TheWayForward} outlines a speculative proof that Splay is
approximately monotone. The appendices formalize relevant folklore.

\section{Preliminaries}\label{sec:Preliminaries}

A \emph{binary tree} $T$ comprises a finite set of \emph{nodes}, with one node
designated to be the \emph{root}. All nodes have a \emph{left} and a
\emph{right} \emph{child} pointer, each leading to a different node. Either or
both children may be \emph{missing}; a missing child is denoted by $\Null$.
Every node in $T$, save for the root, has a single \emph{parent} node of which
it is a child. (The root has no parent.) Each pairing of a node with its parent
is an \emph{edge} in $T$. The \emph{size} of $T$ is the number of nodes it
contains, and is denoted $|T|$. There is a unique path from $\Root(T)$ to every
other node $x$ in $T$, called the \emph{access path} for $x$ in $T$. If $x$ is
on the access path for $y$ then $x$ is an \emph{ancestor} of $y$, and $y$ is a
\emph{descendant} of $x$. If these two nodes are distinct then $x$ is a
\emph{strict} ancestor of $y$ and $y$ is a \emph{strict} descendant of $x$.
(Every node is an ancestor and a descendant of itself.) The subgraph comprising
all descendants of $x$ is called the \emph{subtree rooted at $x$}. Nodes thus
have \emph{left} and \emph{right subtrees} rooted respectively at their left
and right children. (Subtrees are \emph{empty} for $\Null$ children.) The
\emph{depth} of $x$, denoted $\Depth_T(x)$, is the length, in nodes, of its
access path. A \emph{rooted hull} in $T$ is a connected subgraph of $T$ that
includes the root. A rooted hull is itself a binary tree.

In a \emph{binary search tree}, every node has a unique \emph{key}, and the
tree satisfies the \emph{symmetric order} condition: every node's key is
greater than those in its left subtree and smaller than those in its right
subtree. The binary search tree derives its name from how its structure enables
finding keys. To find a requested key, initialize the current node to be the
root. While the current node is not $\Null$ and does not contain the requested
key, replace the current node by its left or right child depending on whether
requested key is smaller or larger than the key in the current node,
respectively. The search returns the last current node, which contains the
requested key if said key is in the tree and otherwise $\Null$. The \emph{left
spine} of $T$ is the access path to the smallest key in $T$, and the
\emph{right spine} of $T$ is the access path to the largest key in $T$. (The
spines of the empty tree are empty.) The left and right spines consist entirely
of left and right pointers, respectively. A tree is \emph{flat} if every node
is on the left or right spine. To keep our presentation simple, we assume that
a key and the node containing it can be used interchangeably in binary
comparisons.

We denote by $|X|$ the length of a finite sequence $X$. The symbol ``$\oplus$''
denotes sequence concatenation. (We sometimes write $X_1\oplus\cdots\oplus X_m$
as $\bigoplus_{i=1}^m X_i$.) The \emph{postorder} of the empty tree is the
empty sequence, and the postorder of binary search tree $T$ whose root $r$ has
left and right subtrees $L$ and $R$ is $\Postorder(L) \oplus \Postorder(R)
\oplus (r)$. The \emph{index} of a given key in $T$ is the number of keys in
$T$ that are less than or equal to the given key. The function mapping each key
in $T$ to its index is the \emph{index map} for $T$, and the inverse of this
function is the \emph{reverse index}. The index map of a finite totally ordered
set is defined analogously. Two binary search trees are \emph{isomorphic} if
relabelling the keys in each tree to their respective indices produces trees
with the same postorder.

To perform a \emph{transformation} on tree $T$, first select an arbitrary
rooted hull $Q$ in $T$. Then reshape $Q$ into any other binary search tree $Q'$
containing the same set of keys. We refer to $Q'$ as a \emph{transition tree}.
To complete the operation, form the \emph{after-tree} $T'$ by substituting $Q'$
for $Q$ in $T$, re-attaching the subtrees of $Q$ to $Q'$ in the manner uniquely
prescribed by the symmetric order.

\begin{figure}
\centering
\includegraphics[width=0.7\columnwidth]{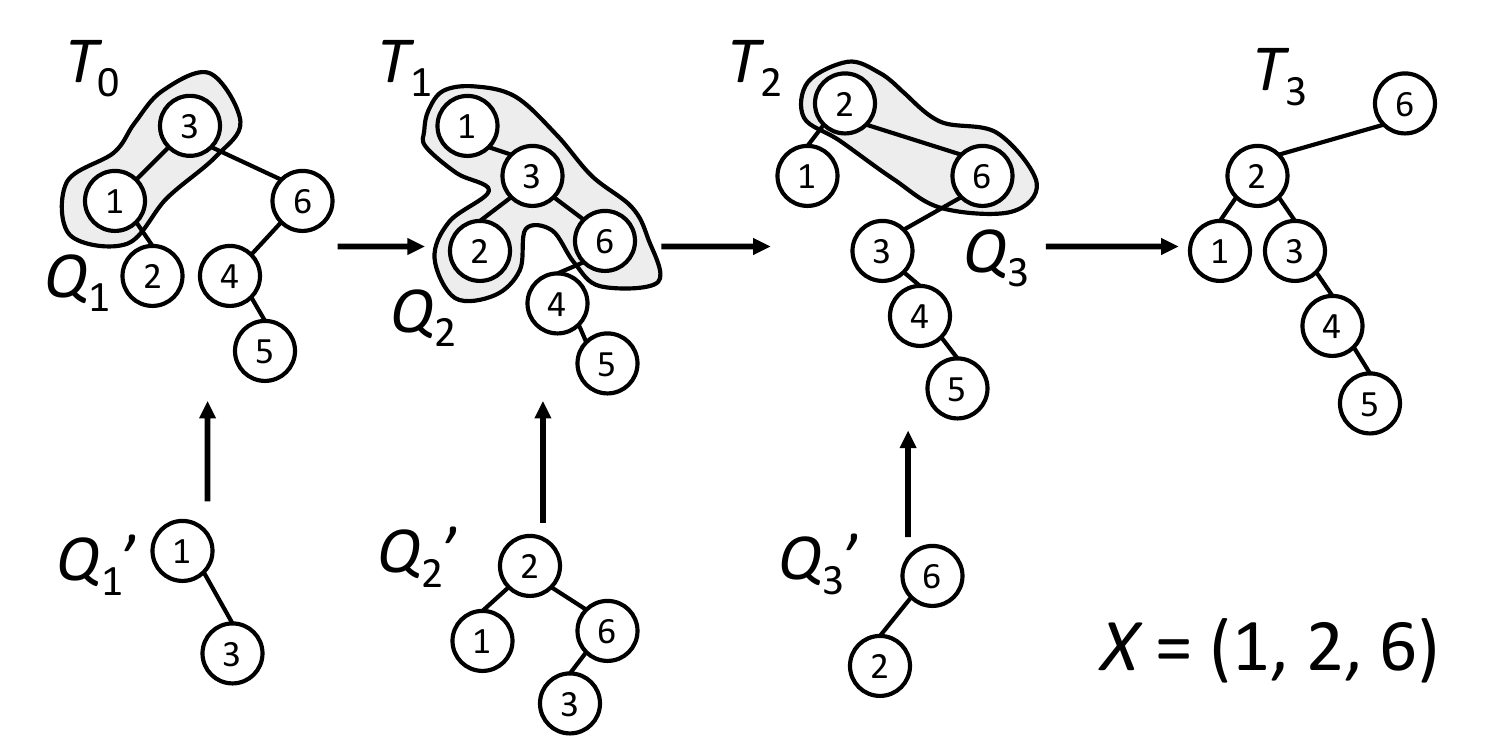}
\caption{Example of an execution for a binary search tree instance. The total cost is eight.}
\label{fig:TransitionTrees}
\end{figure}

An \emph{instance} of a binary search tree optimization problem comprises a
sequence $X=(x_1,\dots,x_m)$ of requested keys and an initial tree $T$
containing these keys. An \emph{execution} $E$ for this instance comprises a
sequence of rooted hulls $Q_1,\dots,Q_m$, a sequence of transition trees
$Q'_1,\dots,Q'_m$, and a sequence of after-trees $T_1,\dots,T_m$. For $1\le
i\le m$, $Q_i$ is a rooted hull in $T_{i-1}$, $Q'_i$ is a binary search tree
with the same keys as $Q_i$ such that $x_i=\Root(Q'_i)$, and $T_i$ results from
substituting $Q'_i$ for $Q_i$ in $T_{i-1}$, where $T_0=T$. (We refer to $T_m$
as the execution's \emph{final tree}.) The \emph{cost} of $E$ is $\sum_{i=1}^m
|Q'_i|$. At least one execution for $X$ starting from $T$ has minimum, or
\emph{optimum} cost, and we denote this cost by $\OPT(X,T)$.
Figure~\ref{fig:TransitionTrees} shows an instance and a corresponding
execution.

An execution's rooted hull and after-tree for a given request are uniquely
determined by the previous after-tree and the request's transition tree. Thus
we shall occasionally denote an execution by its sequence of transition trees.
Defining the cost of an execution as the sum of the transition tree sizes
captures the notion of paying for restructuring: fewer operations are required
to substitute a smaller tree. Each rooted hull contains the access path, which
accounts for the cost of searching. We describe instances that include
insertions and deletions in Section~\ref{sec:Mutation}.

Unless otherwise implied, we may assume without loss of generality that every
node in the initial tree $T$ has a descendant in $T$ whose key is requested in
$X$~\cite[Theorem~43]{LANDSCAPE}, in which case every key in $T$ appears in at
least one transition tree and $\OPT(X,T)\ge|T|$. Similarly, $\OPT(X,T)\ge|X|$
since every execution produces at least one transition tree per request.
Furthermore, since an optimal algorithm can reshape the entire initial tree on
the first request, $\OPT(X,T)\le\OPT(X,T')+|T|$ for any pair of valid binary
search trees $T$ and $T'$ for request sequence $X$ with the same keys.
Therefore, it makes little difference to optimal executions whether the initial
tree is left specified or unspecified, and many authors do not distinguish
between $\OPT(X,T)$ and $\OPTMin(X)=\min_{T\text{ for }X}\OPT(X,T)$. However,
the initial tree can, potentially, have a significant impact on algorithmic
behavior. Thus, we require instances to specify an initial tree. We discuss
this further in Section~\ref{sec:StartupOverhead}.

A binary search tree \emph{algorithm} $\Algorithm$ maps each instance to an
execution of the instance. We denote the cost of this execution by
$\Cost_\Algorithm(X,T)$. We say $\Algorithm$ is \emph{dynamically optimal} if
there is some constant $c\ge 1$ so that $\Cost_\Algorithm(X,T) \le c\OPT(X,T)$
for all request sequences $X$ and all corresponding initial trees $T$. (Other
terms include ``constant-competitive'' and ``instance optimal.'')

\begin{figure}
\centering
\includegraphics[width=0.6\columnwidth]{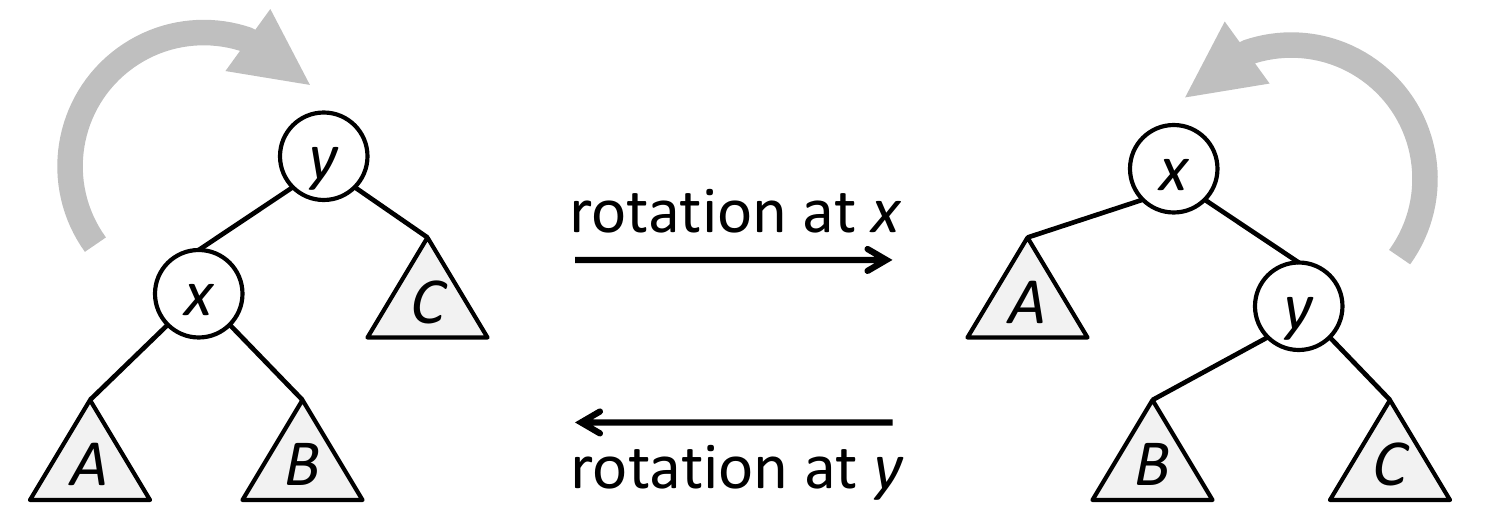}
\caption{Rotation at node $x$ with parent $y$, and reversing the effect by rotating at $y$. Triangles denote subtrees.}
\label{fig:Rotation}
\end{figure}

A \emph{rotation} at left child $x$ with parent $y$ in $T$ replaces the subtree
rooted at $y$ with the tree whose root $x$ has right child $y$ such that the
right subtree of $x$ before the rotation becomes the left subtree of $y$
afterward and the left subtree of $x$ and right subtree of $y$ are unchanged.
Figure~\ref{fig:Rotation} depicts this process. We can also identify this
rotation with the edge connecting $x$ to $y$ in $T$. Rotation at a right child
is symmetric, and rotation at the root is undefined. Rotation preserves
symmetric order while changing up to three child pointers in the tree. Sleator
and Tarjan originally measured execution cost by counting rotations. (See
Appendix~\ref{app:RotationalExecution}.)

A \emph{splay} operation begins with a binary search for a key in the tree. Let
$x$ be the node returned by this search. If $x$ is not $\Null$ then the
algorithm repeatedly applies a splay step until $x$ becomes the root. A
splay step has one of three forms. If the parent of $x$ is the root then rotate
at $x$. (This case is always terminal.) Otherwise, if $x$ is a left child and
its parent is a right child, or vice-versa, rotate at $x$ twice. Otherwise,
rotate at the parent of $x$, and then rotate at $x$. Sleator and Tarjan
assigned the respective names \emph{zig}, \emph{zig-zag} and \emph{zig-zig} to
these three cases~\cite{SPLAY_JOURNAL}. The series of splay steps that bring
$x$ to the root are collectively called splaying at $x$, or simply splaying
$x$. If $X=(x_1,\dots,x_m)$ is a sequence of requested keys in $T$ then the
cost of splaying $X$ starting from $T$ is $\sum_{i=1}^m \Depth_{T_{i-1}}(x_i)$,
where $T_0=T$ and $T_i$ is the result of splaying $x_i$ in $T_{i-1}$ for $1\le
i \le m$. We will primarily be dealing with the Splay algorithm, so
$\Cost(X,T)$, without subscript, will always refer to the cost of splaying the
keys of $X$ starting from $T$.

While an individual splay path can involve every node in the tree, the mean
cost of a splay operation, averaged over sufficiently many requests, is
logarithmic in the tree's size~\cite[Theorem~1]{SPLAY_JOURNAL}. This
performance is similar to that of balanced binary search trees. What makes
Splay remarkable is that it also takes advantage of latent structure in the
request sequence. The amortized cost per splay operation is logarithmic in the
number of unique keys requested since the previous request for the splayed
key~\cite[Theorem~4]{SPLAY_JOURNAL}. Thus, Splay exploits temporal locality in
the access pattern. Splay simultaneously exploits spatial locality. The
amortized cost of a splay operation is logarithmic in the difference between
successively requested keys' indices in the starting
tree~\cite{DYNAMIC_FINGER_2,DYNAMIC_FINGER_1}. The \emph{dynamic optimality
conjecture} states that Splay is dynamically optimal.

Splay has many generalizations. Subramanian defined a class of algorithms that
reshape a tree in small steps. A set of rules, called a ``template,''
determines which step to take based on the arrangement of nodes in the
immediate vicinity of the currently selected node. Different templates give
rise to different algorithms, and a number of these algorithms have many of the
same properties as Splay~\cite{EXPLANATION_OF_SPLAYING}. Georgakopoulos and
McClurkin~\cite{TEMPLATE_CALCULUS} and later Chalermsook et
al.~\cite{GLOBAL_VIEW} proved further results about related algorithms.
Section~\ref{sec:NaturalAlgorithms} examines a generalization of template
algorithms.

Besides Splay, the main candidate algorithm for optimality is colloquially
known as \emph{Greedy}. Lucas and Munro independently conjectured a version of
the algorithm that arranges keys on the access path according to their soonest
future access times is dynamically optimal~\cite{CANONICAL_FORMS,MUNRO_GREEDY}.
Demaine et al.\ subsequently developed a representation of binary search tree
executions as cartesian coordinate point-sets~\cite{GEOMETRY}. They showed that
in this geometric representation, Greedy executes a new request by uniting its
execution of the previous requests with the minimum set of points needed to
make the new execution satisfy some required properties. Subsequently, many of
the interesting behaviors that first drew attention to Splay have been proved
for Greedy, including exploitation of temporal
locality~\cite{FOX,GUPTA_ACCESS_LEMMA} and spatial
locality~\cite{GREEDY_DYNAMIC_FINGER}, as well as some additional
properties~\cite{PATTERN_AVOIDANCE,SIMPLER_PATTERN_AVOIDANCE}.

Two other algorithms serve primarily demonstrative purposes. ``Tango'' trees
require cost proportional to at most $\lg\lg|T|$ times the optimum cost in
order to execute an instance~\cite{TANGO}, and Iacono describes a
multiplicative weights update method that is optimal so long as a certain class
of binary search tree algorithm contains an optimal member~\cite{IN_PURSUIT}.
Both are difficult to implement.

Wilber derived two lower bounds on the cost of executions for a given
instance~\cite{WILBER}. One of his lower bounds counts the number of
occurrences of certain structural patterns in the request sequence with respect
to a given reference tree. We examine Wilber's other lower bound, which we call
the \emph{crossing bound}, in Section~\ref{sec:CrossingCost}. A third lower
bound, called the ``independent rectangle bound,'' is defined
geometrically~\cite{GEOMETRY}. Research into the relationships among these
bounds is ongoing~\cite{WILBER_SEPARATION,CROSSING_VS_IRB}.

Currently, there is no sub-exponential time algorithm that is known to compute
the cost of an optimum binary search tree execution for an instance to within a
constant factor. Circumstantial evidence indicates that exact computation of
optimal execution cost may be intractable, since a slight generalization of the
problem, in which instances comprise requests for batches of keys, is
NP-Complete~\cite{GEOMETRY}. The theoretical and practical difficulties we
encountered when trying to reason about optimal binary search tree executions
ultimately led us to the present approach, which consciously avoids directly
comparing algorithms with optimal behavior.

\section{Approximate Monotonicity}\label{sec:ApproximateMonotonicity}

How can we prove that Splay is dynamically optimal without knowing what optimum
executions ``look like?'' We approach this question by combining two concepts.
The first starts with a simple observation: in many situations, one intuitively
expects that removing requests from an instance should decrease the cost for
the algorithm to execute it. This may not always be the case, but it is a
reasonable idea to explore. The second idea is to force an algorithm to
simulate executions by feeding it appropriately constructed instances. An
algorithm $\Algorithm$ is \emph{approximately monotone} if there is some
constant $b\ge 1$ so that $\Cost_\Algorithm(Y,T)\le b\Cost_\Algorithm(X,T)$ for
every request sequence $X$, subsequence $Y$, and initial tree $T$. A
\emph{simulation embedding} $\Simulate$ for $\Algorithm$ is a map from
executions to request sequences for which there exists $c\ge 1$ such that
$\Cost_\Algorithm(\Simulate(E),T)$ is at most $c$ times the cost of $E$ and $X$
is a subsequence of $\Simulate(E)$ for all instances $(X,T)$ and corresponding
executions $E$. If such a map exists then $\Algorithm$ is \emph{coercible}.

We add a few clarifying comments on terminology. A subsequence need not be
contiguous. For example, $(1,3,6)$ is a subsequence of $(1,2,3,5,6)$. Also,
every sequence is a subsequence of itself. A real-valued set function $F$ is
monotone if $F(A)\le F(B)$ for all $A\subseteq B$. Approximate monotonicity
relaxes this requirement. (The functions we deal with are sequence-valued, but
the concept is identical.) Our SODA paper referred to approximate monotonicity
as the ``subsequence property''~\cite{A_NEW_PATH}. In this work, we only build
simulation embeddings for binary search tree algorithms. However, the concept
itself seems more general and likely has other applications.

\begin{theorem}\label{thm:OptIsMonotone}
Optimal algorithms are monotone.
\end{theorem}

\begin{proof}
Let $X$ be a sequence of $m$ requests with starting tree $T$, let
$E=(Q'_1,\dots,Q'_m)$ be an optimal execution of this instance with after-trees
$T_1,\dots,T_m$, and form subsequence $Y$ from $X$ by choosing a subset $A$ of
$\Set{1,\dots,m}$ and keeping the requests in $X$ at times in $A$. The requests
at times $\Set{1,\dots,m}\setminus A$ can be partitioned into contiguous blocks
of integers. (For example, if $m=11$ and $A=\Set{3,7,9}$ then the removed time
blocks are $\Set{1,2}$, $\Set{4,5,6}$, $\Set{8}$ and $\Set{10,11}$.) Let
$\alpha$ be the index map for $A$. Define the transition tree sequence
$F=(P'_1,\dots,P'_{|Y|})$ as follows. For $i\in A$, if $i$ is one greater than
the maximal element in a removed time block then set $P'_{\alpha(i)}$ to be the
rooted hull in $T_i$ comprising the union keys in $Q'_i$ with the keys in the
transition trees of $E$ for the requests in said block, and otherwise set
$P'_{\alpha(i)}=Q'_i$. The transition tree sequence $F$ is a valid execution
for $Y$ starting from $T$, and $\sum_{i\in A} |P'_{\alpha(i)}| \le \sum_{1\le
i\le m} |Q'_i|$. Since $E$ is an optimal execution for $X$ starting from $T$,
we conclude $\OPT(Y,T)\le\OPT(X,T)$.
\end{proof}

\begin{theorem}\label{thm:CoercibleMonotoneIsOpt}
A coercible algorithm is dynamically optimal if and only if it is approximately
monotone.
\end{theorem}

\begin{proof}
A simulation embedding can be used to simulate an optimal execution of a given
instance just as well as any other execution. The cost for the algorithm to
execute the simulation is no more than a fixed multiple of the optimal cost for
that instance. The simulation of this optimal execution contains the original
request sequence as a subsequence. If the algorithm is also approximately
monotone, then the cost of executing the original instance will not exceed a
fixed multiple of the simulation's cost and hence of the optimal cost. In the
other direction, if $\Algorithm$ is dynamically optimal then there exists some
constant $c$ for which $\Cost_\Algorithm(Y,T)\le c\OPT(Y,T)\le c\OPT(X,T)$ for
all instances $(X,T)$ and subsequences $Y$ of $X$, where the last inequality
follows from Theorem~\ref{thm:OptIsMonotone}.
\end{proof}

Approximate monotonicity is useful even if an algorithm is not dynamically
optimal. For $n>0$, define the \emph{subsequence overhead} $f(n)$ and
\emph{optimal overhead} $h(n)$ of $\Algorithm$ to be the respective suprema of
$\Cost_\Algorithm(Y,T)/\Cost_\Algorithm(X,T)$ and
$\Cost_\Algorithm(X,T)/\OPT(X,T)$ taken over all instances $(X,T)$ and all
subsequences $Y$ of $X$ for which $|T|=n$.

\begin{theorem}\label{thm:CoercibleOverheadEquivalence}
For all $n>0$, a coercible algorithm's subsequence overhead and optimal
overhead are within a constant factor independent of $n$.
\end{theorem}

\begin{proof}
By Theorem~\ref{thm:OptIsMonotone}, $\Cost_\Algorithm(X,T) \ge \OPT(X,T) \ge
\OPT(Y,T)$ for every instance $(X,T)$ and subsequence $Y$ of $X$, meaning
$\Cost_\Algorithm(Y,T)/\Cost_\Algorithm(X,T) \le
\Cost_\Algorithm(Y,T)/\OPT(Y,T) \le h(|T|)$. If $\Algorithm$ is coercible then
there exists a simulation embedding $\Simulate$ and constant $c$ such that
$\OPT(X,T)\ge \Cost_\Algorithm(\Simulate(E),T)/c$ for every optimal execution
$E$ of request sequence $X$ with starting tree $T$. Therefore
$\Cost_\Algorithm(X,T)/\OPT(X,T) \le
c\Cost_\Algorithm(X,T)/\Cost_\Algorithm(\Simulate(E),T) \le c f(|T|)$. Since
these inequalities hold for all instances with an initial tree of size $n$,
they hold true for the supremum. Thus $1\le h(n)/f(n)\le c$ for all $n$.
\end{proof}

To build simulation embeddings, we employ an algorithm for transforming a
binary search tree $T$ into another binary search tree $T'$ with the same keys
through the application of at most $4|T|$ \emph{restricted} rotations, which
must occur at children or grandchildren of the root. Begin by repeatedly
rotating at the root's left child until all nodes in the left subtree of the
root are on the left spine. Then, repeat the following until the root's left
and right subtrees are respectively left and right spines: repeatedly rotate at
the left child of the root's right child so long as said left child is not
$\Null$, and then rotate at the root's right child. Once the tree is flat,
continually rotate at either the left or the right child of the root until the
tree is the same as that resultant from applying the above flattening procedure
to $T'$. Finally, apply the reverse flattening procedure to recover $T'$.
Cleary and Taback first derived this algorithm using group
theory~\cite{CLEARY_RESTRICTED_ROTATIONS}. Our description is based on Lucas'
presentation~\cite{LUCAS_RESTRICTED_ROTATIONS}.

\begin{theorem}\label{thm:SplayTransforms}
For every pair of binary search trees $T$ and $T'$ of size at least four with
the same keys, there exists a request sequence such that Splay's execution of
the requests starting from $T$ has cost linear in $|T|$ and has final tree $T'$.
\end{theorem}

\begin{figure}
\centering
\includegraphics[width=0.7\columnwidth]{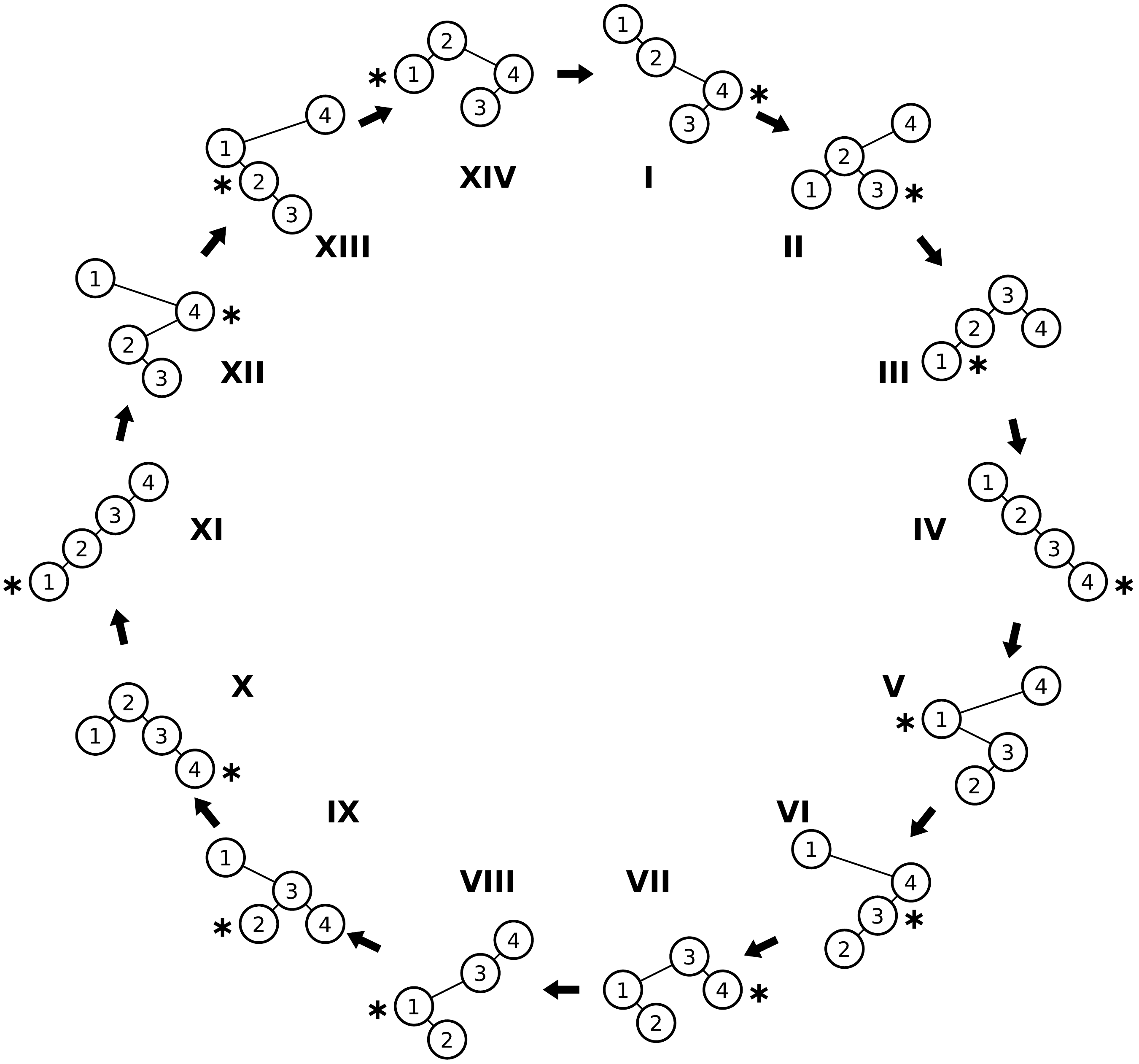}
\caption{Splay at starred nodes to convert one tree into the next.}
\label{fig:HamCycle}
\end{figure}

\begin{proof}
Let $u_1,\dots,u_k$ be the sequence of keys at which Lucas' restricted rotation
algorithm performs rotations in order to transform $T$ into $T'$. Let $T_0=T$
and for $1\le i\le k$ let $q_i$ be the tree of lexicographically smallest
postorder among four-node rooted hulls in $T_{i-1}$ that contain $u_i$. (Using
minimal postorder is just a convention.) Form $q'_i$ by rotating at $u_i$ in
$q_i$ and form $T_i$ by substituting $q'_i$ for $q_i$ in $T_{i-1}$. Form the
key sequence $U_i$ by relabeling the keys in Figure~\ref{fig:HamCycle} via the
reverse index for $q_i$ and recording the sequence of keys in marked nodes on
the path from $q_i$ to $q'_i$, excluding the key marked in $q'_i$. Splaying
$U_i$ starting from $q_i$ results in final tree $q'_i$. The structure of the
subtrees hanging from the path do not affect the transition tree of a splay
operation. Thus, using splay operations to induce a restricted rotation in a
four-node rooted hull in a larger tree $T$ also performs the restricted
rotation in $T$, and the request sequence $V=U_1\oplus\cdots\oplus U_k$ induces
Splay to successively enact the restricted rotations that transform $T$ into
$T'$. Each of these rotations corresponds to at most thirteen requests in $V$.
Every access path in Splay's execution of $V$ starting from $T$ has length at
most four. Since $k<4|T|$, the total cost of this execution is at most $208|T|$.
\end{proof}

\begin{theorem}\label{thm:SplayIsCoercible}
Splay is dynamically optimal if and only if it is approximately monotone.
\end{theorem}

\begin{proof}
We prove Splay is coercible. Let $E$ be an execution for $X=(x_1,\dots,x_m)$
starting from $T$ comprising rooted hulls $Q_1,\dots,Q_m$, transition trees
$Q'_1,\dots,Q'_m$, and after-trees $T_1,\dots,T_m$. For initial trees of size
three or less set $\Simulate(E)=X$. Otherwise, let $\Simulate(E) =
V_1\oplus\cdots\oplus V_m$ where, for $1\le i\le m$, $V_i$ is the request
sequence constructed in Theorem~\ref{thm:SplayTransforms} for inducing
Splay to transform $Q_i$ into $Q'_i$. (If $Q_i=Q'_i$ then $V_i$ is the
singleton request sequence whose sole term is $x_i$.) Splaying $V_i$ starting
from $T_{i-1}$ induces a substitution of $Q'_i$ for $Q_i$ in $T_{i-1}$ to form
$T_i$, where $T_0=T$. A splay operation always places the requested key as the
root of the after-tree. Since $x_i$ is the root of $Q'_i$ it must be the last
key in $V_i$. Therefore, $X$ is a subsequence of $\Simulate(E)$, and by
Theorem~\ref{thm:SplayTransforms} $\Cost(\Simulate(E), T) \le
208(|Q'_1|+\cdots+|Q'_m|)$. Hence, $\Simulate$ is a simulation embedding for
Splay and Splay is coercible. Apply Theorem~\ref{thm:CoercibleMonotoneIsOpt}.
\end{proof}

Splay is only \emph{approximately} monotone. Let $T$ be a left spine with
integer keys $1$ to $2^k-1$ for $k>3$, let $Y=\bigoplus_{i=0}^{k-1}(2^i)$ be
the geometric sequence ascending in powers of two, let $Y'$ be the reversal of
$Y$ and let $X=Y'\oplus Y$. Splaying the first half of $X$ efficiently brings
the requested keys close to the root and ensures $\Cost(X,T)=2^k+4k-5$.
Meanwhile, splaying each request in $Y$ only halves the depth of the next
requested key, so $\Cost(Y,T)=2^{k+1}-3$. Hence the limit of Splay's
subsequence overhead, as tree size increases, is at least two.

Our simulation embedding is designed for minimalism. A more careful analysis
can reduce the constant factor. Two prior works construct simulation embeddings
for binary search tree algorithms. Harmon builds a simulation embedding for the
geometric version of Greedy~\cite[Chapter~2.3.4]{HARMON_THESIS}, while Russo's
simulation embedding for Splay uses rotation-based executions and
potential-based analysis~\cite{RUSSO}. Neither work treats simulation
embeddings as mathematical objects in their own right. Subsequent to the
publication of our SODA paper~\cite{A_NEW_PATH}, Chalermsook and Jiamjitrak
constructed simulation embeddings for a class of template-like algorithms using
potential-based methods~\cite{NEW_SIMULATION_EMBEDDINGS}. Reddmann examines
several algorithms' competitive overheads numerically~\cite{COMPETITIVE_RATIOS}.

\section{Startup Overhead}\label{sec:StartupOverhead}

In principle, an algorithm may need to execute many requests in order to bring
a poorly structured initial tree into a good state before it can behave
optimally. Formally, an algorithm $\Algorithm$ is \emph{eventually optimal} if
there exists a positive constant $b$ and \emph{startup overhead} $g$ mapping
starting trees to integers such that $\Cost_\Algorithm(X,T)\le b\OPT(X,T)+g(T)$
for all request sequences $X$ and corresponding initial trees $T$. Similarly,
if $\Cost_\Algorithm(Y,T)\le b\Cost_\Algorithm(X,T)+g(T)$ for all instances
$(X,T)$ and subsequences $Y$ of $X$ then $\Algorithm$ is \emph{eventually
monotone}. (Eventual optimality implies eventual monotonicity.) 

Some works do not distinguish between eventual and dynamic optimality, but
Sleator and Tarjan were more optimistic. They made no allowance for startup
overhead in their original statement of the dynamic optimality
conjecture~\cite{SPLAY_JOURNAL}. As we shall show, their optimism was
well-placed: if Splay is eventually optimal then it is dynamically optimal. Our
proof bounds startup overhead by averaging execution cost over many repetitions
of a request sequence. A \emph{repeater} $\Repeat$ for algorithm $\Algorithm$
is a mapping from integer-instance pairs to request sequences for which there
exists positive constants $a$ and $c$ such that $k\Cost_\Algorithm(X, T) \le
a\Cost_\Algorithm(\Repeat(k, X, T), T)$ and $\OPT(\Repeat(k, X, T), T) \le
ck\OPT(X, T)$ for all $k>0$, request sequences $X$, and starting trees $T$. If
a repeater exists, then $\Algorithm$ is \emph{repeatable}.

\begin{theorem}\label{thm:EventualOptimality}
Eventually optimal repeatable binary search tree algorithms are dynamically
optimal.
\end{theorem}

\begin{proof}
Repeatability and eventual optimality imply positive constants $a$, $b$ and $c$
and startup overhead $g$ such that $k\Cost_\Algorithm(X,T) \le a
\Cost_\Algorithm(\Repeat(k, X, T), T) \le ab \OPT(\Repeat(k, X, T), T) + a g(T)
\le (abc)k\OPT(X, T) + a g(T)$ for all request sequences $X$, starting trees
$T$ and $k>0$. Choose $k\ge g(T)/\OPT(X, T)$ to absorb the overhead and obtain
$\Cost_\Algorithm(X,T)\le a(bc+1)\OPT(X,T)$.
\end{proof}

\begin{theorem}\label{thm:EventualMonotonicity}
Eventually monotone repeatable coercible algorithms are dynamically optimal.
\end{theorem}

\begin{proof}
These properties imply there exists a simulation embedding $\Simulate$,
constant $b$ and startup overhead $g$ for which $\Cost_\Algorithm(X,T) \le
b\Cost_\Algorithm(\Simulate(E),T)+g(T) \le b \OPT(X,T)+g(T)$ for all instances
$(X,T)$ and corresponding optimal executions $E$. Apply
Theorem~\ref{thm:EventualOptimality}.
\end{proof}

\begin{theorem}\label{thm:SplayIsRepeatable}
If Splay is eventually monotone then it is dynamically optimal.
\end{theorem}

\begin{proof}
We show Splay is repeatable. Let $X$ be a request sequence with initial tree
$T$. Let $V$ be the final tree in Splay's execution of $X$ starting from $T$,
and define the extended sequence $U=X\oplus W$, where $W$ is the sequence
described in Theorem~\ref{thm:SplayTransforms} that induces Splay to
transform $V$ into $T$. (If $|T|<4$ or $V=T$ then $W=\varnothing$.) Denote by
$k*U$ the sequence $U$ repeated $k$ times. Since $U$ merely consists of
requests appended to $X$, $\Cost(X,T)\le \Cost(U,T)$. The final tree in Splay's
execution of $U$ starting from $T$ is again $T$, so each repetition has
identical after-trees, and $\Cost(k*U,T)=k\Cost(U,T)$. Thus,
$k\Cost(X,T)\le\Cost(k*U,T)$.

It remains to bound the optimal cost. If $X\ne\varnothing$ let $A=(T')\oplus B$
where $T'$ is the after-tree for the first request in some optimal execution
$E$ for $X$ starting from $T$ and $B$ is the sequence of transition trees in
$E$ for the remaining requests in $X$, otherwise let $B=\varnothing$.
Similarly, if $W\ne\varnothing$ let $C=(V')\oplus D$ where $V'$ is the first
after-tree in Splay's execution $F$ for $W$ starting from $V$ and $D$ is the
sequence of transition trees in $F$ for the remaining requests in $W$,
otherwise let $C=\varnothing$. The sequence of transition trees $G=A\oplus C$
is an execution of $U$ starting from $T$. The transition trees in $A$ have
total size at most $|T|+\OPT(X,T)$, and by
Theorem~\ref{thm:SplayTransforms} the transition trees in $C$ have total
size at most $209|T|$. Since we can absorb initial tree size into optimal cost,
$\OPT(U,T)\le 211\OPT(X,T)$. Finally, $k*G$ is an execution for $k*U$ starting
from $T$, meaning $\OPT(k*U,T)\le 211k\OPT(X,T)$, and $\Repeat(k,X,T)=k*U$ is a
repeater for Splay. Apply Theorems~\ref{thm:EventualMonotonicity}
and~\ref{thm:SplayIsCoercible}.
\end{proof}

Our SODA paper established the contrapositive of
Theorem~\ref{thm:SplayIsRepeatable} by repeating hypothetical instances on
which Splay is non-optimal in order to contradict any presumed nontrivial
startup overhead~\cite{A_NEW_PATH}. Kurt Mehlhorn kindly supplied us with an
outline of the above version of the proof after he reviewed our manuscript.

\section{Mutation}\label{sec:Mutation}

Monotonicity has no clear analog for algorithms that can handle requests to add
and remove keys from the tree. We work around this obstacle by representing
these operations using executions of instances that lack such requests. A
\emph{reduction} $\Reduce$ from algorithm $\Algorithm$ in execution model
$\Model$ to algorithm $\Algorithm'$ in execution model $\Model'$ is a map from
instances in $\Model$ to instances in $\Model'$ for which there exists positive
constants $a$ and $c$ such that $\Cost_\Algorithm(\Instance)\le
a\Cost_{\Algorithm'}(\Reduce(\Instance))$ and
$\OPT_{\Model'}(\Reduce(\Instance))\le c\OPT_\Model(\Instance)$ for all
$\Instance\in\Model$. If such a reduction exists we say $\Algorithm$
\emph{reduces} to $\Algorithm'$.

\begin{theorem}\label{thm:Reduction}
If $\Algorithm$ reduces to $\Algorithm'$ and $\Algorithm'$ is
dynamically optimal then $\Algorithm$ is dynamically optimal.
\end{theorem}

\begin{proof}
The reduction to an optimal algorithm $\Algorithm'$ implies the existence of
constants $a$, $b$ and $c$ such that $\Cost_\Algorithm(\Instance) \le
a\Cost_{\Algorithm'}(\Reduce(\Instance)) \le
ab\OPT_{\Model'}(\Reduce(\Instance)) \le abc\OPT_\Model(\Instance)$ for all
$\Instance \in \Model$.
\end{proof}

Our reduction employs the following terminology. To \emph{augment} a binary
search tree $T$ with a new key $k$, first do a search for $k$ in $T$. When the
search reaches a missing node, replace this node with a new node containing the
key $k$. Augmenting an empty tree makes $k$ the root key. (This process is
sometimes called ``leaf insertion.'') The \emph{successor} of $k$ in $T$ is the
smallest key in $T$ that is greater than $k$. If no such key is present the
successor is $\Null$. The \emph{predecessor} is defined symmetrically. The
predecessor and successor are the \emph{neighbors} of $k$ in $T$, and the
\emph{neighborhood} of $k$ is the set comprising $k$ and those of its neighbors
that are not missing. If $k$ has no children then its \emph{removal} from $T$
is the rooted hull comprising every key in $T$ except for $k$, unless $k$ is
the root, in which case its removal forms the empty tree.

A \emph{mutating} instance comprises a request sequence
$\chi=((r_1,x_1),\dots,(r_m, x_m))$ and an initial tree $T$ where each
requested operation $r_i\in\Set{\Search,\Insert,\Delete}$. An execution $E$ of
this instance comprises a sequence of rooted hulls $Q_1,\dots,Q_m$, transition
trees $Q'_1,\dots,Q'_m$, after-trees $T_1,\dots,T_m$ and $T_0=T$. If
$r_i=\Search$ then $x_i$ must be in $T_{i-1}$ and $Q_i$, $Q'_i$ and $T_i$ obey
the same restrictions as instances without mutation. For $1\le i\le m$, if
$r_i=\Insert$ then $x_i$ must not be in $T_{i-1}$ and $Q_i$, $Q'_i$ and $T_i$
fulfill a request to search for $x_i$ in the augmentation of $T_{i-1}$ with
$x_i$. If $r_i=\Delete$ then $x_i$ must be in $T_{i-1}$, $Q_i$ contains the
neighborhood of $x_i$ in $T_{i-1}$, $Q'_i$ contains the neighbors of $x_i$ in
$T_{i-1}$ as a rooted hull of its left spine so long as at least one neighbor
is not missing, and $T_i$ results from substituting $Q'_i$ for $Q_i$ in
$T_{i-1}$ and then removing $x_i$. The cost of $E$ is $\sum_{i=1}^m|Q'_i|$. We
denote by $\OPTmut(\chi, T)$ the minimum cost among executions for $\chi$
starting from $T$.

Requiring that executions incorporate \emph{both} of a deleted key's neighbors,
when they are present, is essential to our analysis. We are unable to determine
if algorithms that are dynamically optimal among executions in our model of
deletion remain so after removing this requirement. Having stated this caveat,
our nonstandard version of deletion does not change any known upper bound on
the optimum cost of executing a mutating instance, that we are aware of, by
more than a constant factor. We believe our model is sufficiently realistic to
proceed without further concern.

Our extension of Splay inserts by augmenting $T$ with $x$ followed by splaying
$x$ and deletes $x$ from $T$ by successively splaying the keys in the
neighborhood of $x$ in $T$ in increasing order, rotating at the predecessor of
$x$ if it is present, and then removing $x$. Splay's transition tree for the
deletion is the rooted hull comprising the union of keys on the access paths of
these operations in the tree immediately prior to the removal of $x$.

\begin{theorem}\label{thm:SplayIsReducible}
If Splay is eventually monotone for instances without mutation then it is
dynamically optimal for instances with mutation.
\end{theorem}

\begin{proof}
We reduce Splay with mutation to Splay without mutation. Let
$\chi=((r_1,x_1),\dots,(r_m, x_m))$ and $T$ be the request sequence and
starting tree of a mutating instance. Construct a new instance without
mutation, as follows. Let $K_0$ be the set of keys in $T$ and form $S_0$ by
relabelling the keys in $T$ to their respective indices. All nodes in $S_0$ are
unmarked. (A node's marking status merely aids in our construction and has no
effect on algorithmic behavior.) For $1\le i\le m$, let $u_i$ be the key whose
index among the set of keys held by unmarked nodes in $S_{i-1}$ is the same as
the index of the predecessor of $x_i$ in $K_{i-1}$ if a predecessor is present,
otherwise set $u_i=-\infty$. Define $w_i$ analogously for the successor of
$x_i$ if a successor is present, otherwise set $w_i=\infty$.

If $r_i=\Search$ then let $z_i$ be the key whose index among those held by
unmarked nodes in $V_{i-1}$ is the same as the index of $x_i$ in $K_{i-1}$, let
$S_i=S_{i-1}$ and set $Y_i=(z_i)$ and $K_i=K_{i-1}$.

If $r_i=\Insert$ then do the following. If $S_{i-1}$ has marked nodes with keys
strictly between $u_i$ and $w_i$ in symmetric order then set $z_i$ to the most
recently marked among them and form $S_i$ by unmarking $z_i$ in $S_{i-1}$.
Otherwise, form $S_i$ by augmenting $S_{i-1}$ with a new unmarked node whose
key $z_i$ is as follows. If neither $u_i$ nor $w_i$ are finite then $z_i=0$; if
only $u_i$ is finite then $z_i=u_i-1$; if only $w_i$ is finite then
$z_i=w_i+1$; otherwise, $z_i$ is the midpoint between $u_i$ and $w_i$ on the
real line. Set $Y_i=(z_i)$ and $K_i=\Set{x_i}\cup K_{i-1}$.

If $r_i=\Delete$ then define $z_i$ as in the case for search. If at least one
of $u_i$ and $w_i$ is non-finite then form $S_i$ by marking $z_i$ in $S_{i-1}$.
If neither $u_i$ nor $w_i$ is finite set $Y_i=(z_i)$; if only $u_i$ is finite
set $Y_i=(u_i,z_i,u_i)$; otherwise, if only $w_i$ is finite set
$Y_i=(z_i,w_i)$. When both $u_i$ and $w_i$ are finite, proceed as follows.
Define $v_i$ to be the key in the most recently marked among the marked nodes
of $S_{i-1}$ with keys strictly between $z_i$ and $w_i$ if such a node is
present, otherwise $v_i$ is the midpoint between $z_i$ and $w_i$ on the real
line. If $v_i\in S_{i-1}$ then $S_i$ is as in the case when at least one of
$u_i$ and $w_i$ is finite, otherwise form $S_i$ by augmenting $S_{i-1}$ with an
unmarked node holding key $v_i$ and successively marking $v_i$ and then $z_i$.
Set $Y_i=(u_i, z_i, w_i, v_i, w_i, u_i, w_i)$. Splaying the first five of these
requests ensures the keys $\Set{u_i,z_i,v_i,w_i}$ comprise a rooted hull of the
left spine, and splaying the final two requests induces successive rotations at
$z_i$ and $u_i$. Finally, set $K_i=K_{i-1}\setminus\Set{x_i}$.

Define $\Reduce(\chi,T)=(Y_1\oplus\cdots\oplus Y_m,S_m)$.
Figure~\ref{fig:Reduction} depicts an example of this process. Let
$P'_1,\dots,P'_m$ and $L_1,\dots,L_m$ be the transition trees and after-trees
in Splay's execution of $\chi$ starting from $T$. Set $B_0=S_m$ and for $1\le
i\le m$ let $A_i$ be the union of keys in the transition trees of Splay's
execution of $Y_i$ starting from from $B_{i-1}$ and let $B_i$ be the final tree
of this execution. If $S_i$ has unmarked nodes then their keys comprise a
rooted hull in $B_i$ which is isomorphic to $L_i$, and $P'_i$ is isomorphic to
a subgraph of the rooted hull in $B_i$ comprising the keys in $A_i$. Thus,
$\Cost(X,S)\le\Cost\Reduce(\chi,S)$.

\begin{figure}
\centering
\includegraphics[width=0.5\columnwidth]{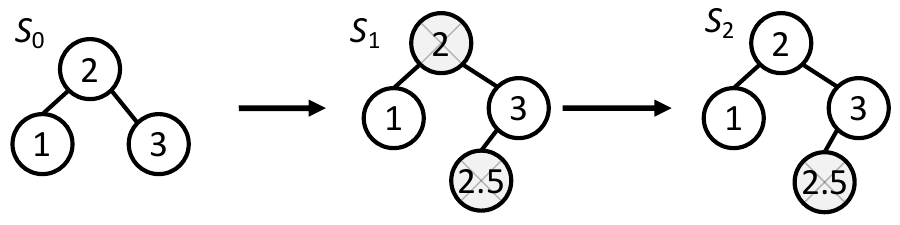}
\caption{After-trees in the construction of $\Reduce(\chi,T)$ where $\chi=((\Delete,12),(\Insert,15))$ and $\Postorder(T)=(2,16,12)$. The request sequence for the reduced instance is $(1,2,3,5/2,3,1,3,2)$.}
\label{fig:Reduction}
\end{figure}

It remains to bound the cost of an optimal execution for the new instance. Let
$Q_1,\dots,Q_m$ be the rooted hulls and $Q'_1,\dots,Q'_m$ be the transition
trees for an optimal execution of $\chi$ starting from $T$, and let $G_0=S_m$.
For $1\le i\le m$, if $r_i\ne\Delete$ or $\infty\in\Set{-u_i,w_i}$ then let
$C_i$ be the rooted hull of $G_{i-1}$ that is isomorphic to $Q_i$ and let
$C'_i$ be the transition tree with the same keys as $C_i$ that is isomorphic to
$Q'_i$. If $r_i=\Delete$ and $\infty\notin\Set{-u_i,w_i}$ then form $C_i$ and
$C'_i$ by respectively augmenting said isomorphic rooted hull and corresponding
transition tree with $v_i$. The first in the transition tree sequence $D'_i$
for the requests in $Y_i$ is the tree that results from splaying the first
requested key of $Y_i$ in $C'_i$. The remaining transition trees in $D'_i$ are
the transition trees of Splay's execution of the remaining requests in $Y_i$
starting from the first tree in $D'_i$. Finally, let $G_i$ be the final
after-tree in the execution of $Y_i$ starting from $G_{i-1}$ whose transition
trees are $D'_i$. The first tree in $D'_i$ has size at most $|Q'_i|+1$, and
$Y_i$ has at most six remaining requests, each served by a transition tree from
$D'_i$ with size at most four, meaning $\sum_{H\in D'_i}|H|\le 26 |Q'_i|$. The
transition tree sequence $\bigoplus^m_{i=1}D'_i$ executes $\Reduce(\chi,T)$.
Thus, $\OPT\Reduce(\chi,T)\le 26\OPTmut(\chi,T)$, and $\Reduce$ is a reduction.
Apply Theorems~\ref{thm:Reduction} and~\ref{thm:SplayIsRepeatable}.
\end{proof}

Theorem~\ref{thm:SplayIsReducible} has an interesting consequence. A \emph{deque
instance} is an initial tree together with a request sequence entirely
comprising $\Push$, $\Pop$, $\Inject$ and $\Eject$ operations which
respectively correspond to inserting a new maximum, deleting the maximum,
inserting a new minimum and deleting the minimum. A binary search tree
algorithm $\Algorithm$ supports deque operations if there is some $c>0$ so that
$\Cost_\Algorithm(D, T)\le c(|D|+|T|)$ for all deque instances $(D,T)$. The
\emph{deque conjecture} states that Splay supports deque operations. Tarjan
proved that Splay supports a limited subset of deque
operations~\cite{SEQUENTIAL_ACCESS}. Sundar placed an inverse-Ackermann upper
bound on the cost of performing general deque operations~\cite{SUNDAR_DEQUE}.
Pettie later tightened this bound~\cite{DAVENPORT_SCHINZEL}. Similar bounds are
known for Greedy~\cite{GREEDY_DEQUE}. We add a new result:

\begin{theorem}\label{thm:Deque}
If Splay is eventually monotone without mutation then it supports deque
operations.
\end{theorem}

\begin{proof}
Let $D=(r_1,\dots,r_m)$ and $T$ be the request sequence and initial tree for a
deque instance, and assume without loss of generality that the starting tree's
keys are integers. We construct after-trees $T_1,\dots,T_m$ for an execution of
this instance, as follows. If $r_i\in\Set{\Inject,\Eject}$ then the root of
$T_i$ is its minimum, assuming $T_i\ne\varnothing$. Otherwise, if
$r_i\in\Set{\Push,\Pop}$ then the root of $T_i$ is its maximum, and if
$r_i=\Pop$ and $\min T_i\ne\max T_i$ then the root's left child is the minimum
key in the tree. The remaining keys $K_i$ in $T_i$ are in a subtree at the
appropriate location in symmetric order with the following structure. The root
$v_i$ of $K_i$ is the largest key less than or equal to the median key in
$K_i$, assuming $K_i\ne\varnothing$. The left and right subtrees of $v_i$ are
respectively right and left spines comprising the keys in
$K_i\setminus\Set{v_i}$ that are less than and greater than $v_i$. The initial
transition tree of this execution has at most $|T|+1$ nodes and the remainder
of the execution can be realized using transition trees each of size at most
eight. Thus, $\OPTmut(D,T)\le 8(|D|+|T|)$. Apply
Theorem~\ref{thm:SplayIsReducible}.
\end{proof}

There are other ways to implement insertion and deletion. Sleator and Tarjan
analyze an extension of Splay which supports mutation operations by using
splits and joins~\cite{SPLAY_JOURNAL}. Tarjan implements $\Push$ and $\Inject$
by inserting at the top of the tree~\cite{SEQUENTIAL_ACCESS}. Zip trees insert
and delete keys starting from the middle of the tree and rearrange descendants
to restore the binary search tree invariants~\cite{ZIP_TREES_JOURNAL}. Common
implementations of deletion in computers replace the deleted node with the node
holding the predecessor or successor of the removed key, a technique originally
devised by Hibbard~\cite{HIBBARD_DELETION}. Our version of deletion originates
from Cole's analysis of the ``dynamic finger'' theorem~\cite{DYNAMIC_FINGER_2}.
Its main advantage is enabling our proof of Theorem~\ref{thm:SplayIsReducible}.
We leave as open problems determining whether Tarjan's implementation of deque
operations or algorithms that use Hibbard's variant of deletion are reducible
to algorithms in our model of mutation.

\section{Natural Algorithms}\label{sec:NaturalAlgorithms}

Our results readily generalize. We say an algorithm is \emph{natural} if the
rooted hulls of its executions are always the access paths for the requested
keys and the transition trees for isomorphic rooted hulls are isomorphic. Splay
is a natural algorithm. We incorporate mutation in the same way as for Splay. A
natural algorithm inserts by searching in the augmented tree and deletes by
successively searching in increasing order for all in the neighborhood of the
deleted key, then rotating at the predecessor if present, followed by removing
the key. To construct the \emph{transition digraph} $\Graph_n(\Algorithm)$ for
natural algorithm $\Algorithm$, assign a vertex to every binary search tree
with keys $\Set{1,\dots,n}$, and for every $T\in\Graph_n(\Algorithm)$ and $x\in
T$ add an arc from $T$ to the result of executing a search for $x$ in $T$ with
$\Algorithm$.

\begin{theorem}\label{thm:NaturalOptimality}
A natural algorithm whose transition digraph is strongly connected for binary
search trees of some size at least three is dynamically optimal for instances
with mutation if and only if it is eventually monotone for instances without
mutation.
\end{theorem}

\begin{proof}
Let $\Algorithm$ be a natural algorithm and let $N$ be the smallest integer
greater than two for which $\Graph_N(\Algorithm)$ is strongly connected. Choose
a map $P$ from each pair $q,q'\in\Graph_N(\Algorithm)$ to some request sequence
whose execution by $\Algorithm$ starting from $q$ has a sequence of after-trees
which, when prefixed by $q$, comprises a directed path of minimal length
connecting $q$ to $q'$ in $\Graph_N(\Algorithm)$. Let $T'$ be the after-tree of
transforming $T$ with rooted hull $Q$ and transition tree $Q'$. If $Q=Q'$ or
$|T|<N$ then define $C_T(Q,Q')=\Root(Q')$. Otherwise, choose rooted hulls
$q_1,\dots,q_k$ and transition trees $q'_1,\dots,q'_k$ for enacting restricted
rotations in an identical manner to Theorem~\ref{thm:SplayTransforms}, except
that each transition tree has size $N$, rather than size four, and set
$C_T(Q,Q')=\bigoplus_{i=1}^k P(q_i,q'_i)$. The output of $P$ contains at most
one request per vertex in $\Graph_N(\Algorithm)$. There are $(2N)!/(N!(N+1)!)$
binary trees with keys $\Set{1,\dots,N}$~\cite{CATALAN_NUMBERS}. (This is the
Catalan number for $N$.) Each access path contains at most every node in the
tree. Thus, the cost for $\Algorithm$ to execute $C_T(Q,Q')$ starting from $T$
is at most $4(2N)!/((N+1)!(N-1)!)|Q'|$, which is linear in the transition
tree's size. This execution's final tree is $T'$ whenever $|T|\ge N$.

The map $\Simulate(E)=\bigoplus_{i=1}^m C_T(Q_i,Q'_i)$, where $Q_1,\dots,Q_m$
and $Q'_1,\dots,Q'_m$ are the rooted hulls and transition trees of some
execution $E$ for $X$ starting from $T$, is a simulation embedding for
$\Algorithm$. Similarly, $\Repeat(k,X,T)=k*(X\oplus C_T(T,V))$, where $V$ is
the final tree in the execution of $X$ starting from $T$ by $\Algorithm$, is a
repeater for $\Algorithm$. Thus, by Theorem~\ref{thm:EventualMonotonicity}, if
$\Algorithm$ is eventually monotone then it is dynamically optimal for
instances without mutation. To reduce how $\Algorithm$ executes mutation
requests to its behavior when executing non-mutating instances, modify how the
reduction in Theorem~\ref{thm:SplayIsReducible} handles requests to delete keys
with both predecessors and successors. Instead of requesting a single auxiliary
key, as in the case for Splay, add requests for $N-3$ auxiliary keys,
augmenting the initial tree with unmarked nodes as necessary. Then, replace the
requests that induce Splay to perform the relevant rotations along the left
spine with the output of $P$. Necessity of eventual monotonicity follows from
Theorem~\ref{thm:OptIsMonotone}.
\end{proof}

A strongly connected transition digraph is not necessary for a natural
algorithm to be coercible. An example of this is a variant of Splay that
carries out restructuring operations in tandem with the binary search for the
requested key, eliminating the need for parent pointers, call stacks or
threaded nodes~\cite{SPLAY_JOURNAL}. A \emph{top-down-splay} operation for $x$
in $T$ begins by initializing a pointer $l$ to a childless node whose key is
$-\infty$, a pointer $r$ to a childless node with key $\infty$, and a pointer
$t$ to the root of $T$. It uses left and right \emph{linking} steps. Linking
left replaces the left subtree of $r$ with $t$, redirects $r$ to point to the
target of $t$, and redirects $t$ to point to its target's left child. Linking
right is symmetric. The operation repeats the following process until $t=x$.
Suppose without loss of generality that $x$ is in the subtree rooted at the
left child $y$ of $t$. (The other case is symmetric.) If $y=x$ then execute a
right link. (This case is terminal.) Otherwise, if $x<y$ then rotate at $y$,
redirect $t$ to point to $y$, and execute a right linking operation. Otherwise,
execute a right link followed by a left link operation. Once $t=x$, the
operation completes by replacing the right subtree of $l$ with the left subtree
of $t$ followed by replacing the left subtree of $t$ with the right subtree of
$-\infty$, and doing symmetrically with $r$, the left subtree of $\infty$ and
the right subtree of $t$. M\"{a}kinen compares Splay to its top-down variant in
detail~\cite{TOP_DOWN_SPLAY}.

\begin{theorem}\label{thm:NoTopDownSplayTransform}
Top-Down Splay's transition digraph is not strongly connected for binary search
trees of any size greater than two.
\end{theorem}

\begin{proof}
Let $S$ be a binary search tree of size at least three whose root $a=\min S$
has right child $z=\max S$, let $Q$ be the left subtree of $z$ in $S$ and let
$T=\TopDownSplay(T,x)$ for some $x\in Q$. We show no request sequence induces
Top-Down Splay to restore $T$ to $S$. Suppose, for the sake of contradiction,
that such a sequence exists. Because $\Root(S)=a$, the last key requested in
any such sequence must be $a$. Since $\TopDownSplay(T,a)\ne S$, there must be
at least one preceding request for a different key. Let $y\ne a$ be penultimate
key in this sequence and let $R$ be after-tree corresponding to this request,
so that $\Root(R)=y$. We demonstrate $\TopDownSplay(R,a)\ne S$.

Suppose first that $y\in Q$. Because $a<y<z$, the left and right subtrees of
$y$ in $R$ respectively contain $a$ and $z$. Thus, $z$ is not on the access
path for $a$ in $R$. Because $y$ is the largest key on the access path to $a$
in $R$ and Top-Down Splay is a natural algorithm, the access path to $y$ in
$\TopDownSplay(R,a)$ is a right spine rooted at $a$, and $y$ is an ancestor of
$z$ in this after-tree. This is incompatible with $z$ being the right child
$a$, as is the case in $S$. Thus, $y=z$. Since $a$ is the smallest key, it is
on the left spine. Direct computation shows $\TopDownSplay(R,a)\ne S$ when $a$
has depth three or four, and a simple induction establishes the same for a left
spine of any greater length. Hence, there is no path from $T$ to $S$ in
Top-Down Splay's transition digraph, and this transition digraph is not
strongly connected.
\end{proof}

\begin{figure}
\centering
\includegraphics[width=0.2\columnwidth]{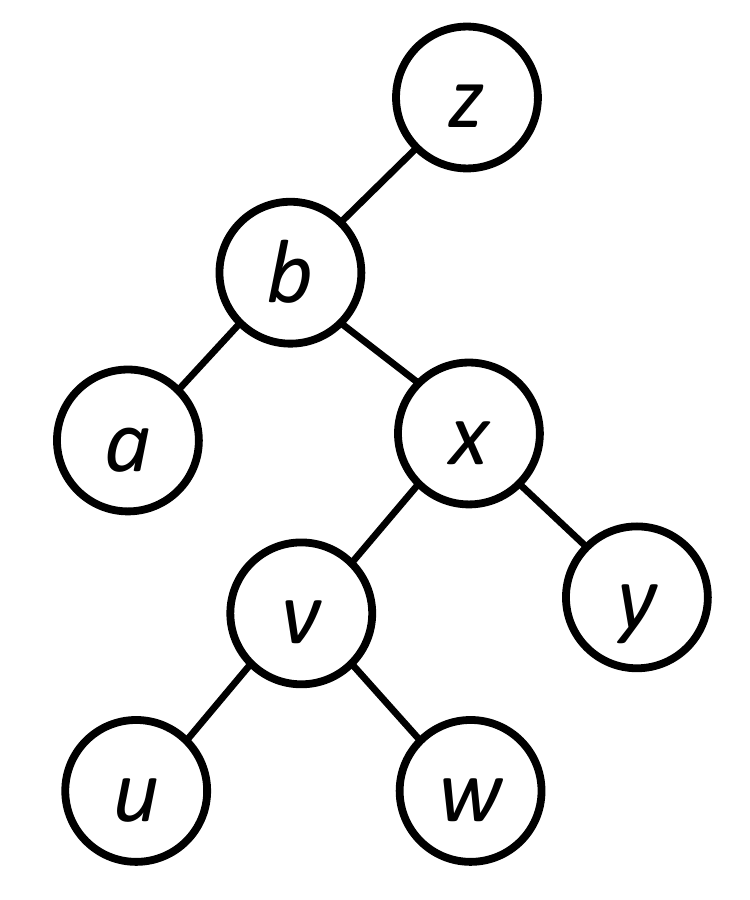}
\caption{Schematic of the top portion of some $\hat{T}_i$. To induce a rotation at $u$, perform top-down splays at $(a,u,a,z)$. To induce a rotation at $v$, top-down splay the sequence $(v,a,z)$. To induce a rotation at $w$, top-down splay at $(a,w,a,z)$. To induce a rotation at $y$, top-down splay at $(y,a,z)$.}
\label{fig:TopDownRestrictedRotations}
\end{figure}

\begin{theorem}\label{thm:TopDownSplayIsCoercible}
If Top-Down Splay is approximately monotone then it is dynamically optimal.
\end{theorem}

\begin{proof}
Let $T$ be a nonempty binary search tree, let $a=\min T$, let $z=\max T$, and
let $b$ be the successor of $a$ in $T$ if the successor is present, otherwise
$b=a$. Define the map $H$ that transforms any binary search tree $S$ with the
same keys as $T$ in the following way. Form $U$ by replacing the left subtree
of the parent of $a$ in $S$ with the right subtree of $a$ in $S$ if the parent
is present, otherwise set $U=S$. Form $V$ from $U$ by doing similar with $b$,
and form $W$ from $V$ by replacing the right subtree of the parent of $z$ with
the left subtree of $z$ in $V$, otherwise $W=V$. The tree $H(S)$ is the binary
search tree whose left spine comprises $\Set{a,b,z}$ such that $W$ is the right
subtree of $b$ in $H(S)$.

Let $E$ be an execution for $X=(x_1,\dots,x_m)$ starting from $T$ with rooted
hulls $Q_1,\dots,Q_m$, transition trees $Q'_1,\dots,Q'_m$ and after-trees
$T_1,\dots,T_m$ and let $T_0=T$. Let $A=(z,b,a,z)$, let $\hat{T}_{-1}$ be the
final tree in Top-Down Splay's execution of let $A$ starting from
$\hat{T}_{-1}$, let $Q_0=\hat{T}_{-1}$, let $Q'_0=T$, let $T_0=T$ and let
$x_0=z$. For $0\le i\le m$ let $\hat{T}_i=H(T_i)$, let $\hat{Q}_i$ and
$\hat{Q}'_i$ respectively be the smallest rooted hulls in $\hat{T}_{i-1}$ and
$\hat{T}_i$ containing the keys in $Q_i\cup\Set{a}$, and let $Y_i$ be the
sequence of keys determined by the restricted rotation algorithm for
transforming the right subtree of $b$ in $\hat{Q}_i$ into its right subtree in
$\hat{Q}'_i$. Form $Z_i$ by replacing each key in $Y_i$ with the corresponding
keys determined by Figure~\ref{fig:TopDownRestrictedRotations}, and then
appending the requests $(x_i,a,z)$. Top-Down Splay's execution of $A$ has cost
proportional to at most $|T|$, which can be absorbed into the cost of $E$. The
sequence $Z_i$ induces Top-Down Splay to transform $\hat{Q}_i$ into
$\hat{Q}'_i$. Since $|\hat{Q}'_i|\le|Q'_i|+3$, the transformation's cost is
proportional to $|Q'_i|$. Thus, the map $\Simulate(E)=A\oplus \bigoplus_{i=0}^m
Z_i$ is a simulation embedding for Top-Down Splay. Apply
Theorem~\ref{thm:CoercibleMonotoneIsOpt}.
\end{proof}

While the rooted hulls of Greedy's executions are access paths, its transition
trees are determined by which keys are in surrounding requests, meaning Greedy
is not a natural algorithm. Lucas conjectured that restricting executions'
rooted hulls to the access path does not increase optimal cost by more than a
fixed multiple~\cite{CANONICAL_FORMS}. This conjecture remains open. Kozma
catalogues related open questions about the relative power of classes of binary
search tree algorithms subject to various restrictions~\cite{KOZMA_THESIS}.
Splay is in the most restrictive of these classes, meaning dynamic optimality
would imply they are all equivalent up to constant factors.

\section{Crossing Cost}\label{sec:CrossingCost}

Binary search trees facilitate efficient search by arranging keys into many
short access paths comprising children of alternating direction, and an
algorithm's efficiency depends critically on how it utilizes these
arrangements. Consider the subtree $P$ of a binary search tree $T$ comprising
the access path for a node $x$ in $T$. The \emph{crossing nodes} for $x$ in $T$
comprise $x$, the root of $T$, and the nodes in $P$ that are either left
children with a right child on $P$ or right children with a left child on $P$.
We refer to the number of crossing nodes for $x$ as its \emph{crossing depth}
in $T$, denoted $\Level_T(x)$. The \emph{bookkeeping nodes} are the
non-crossing nodes on $P$. The \emph{crossing cost} of execution $E$ with
after-trees $T_1,\dots,T_m$ for $X=(x_1,\dots,x_m)$ starting from $T$ is
$\sum_{i=1}^m \Level_{T_{i-1}}(x_i)$ where $T_0=T$, and the execution's
\emph{bookkeeping cost} is $\sum_{i=1}^m
(\Depth_{T_{i-1}}(x_i)-\Level_{T_{i-1}}(x_i))$.

To perform a \emph{move-to-root} operation, repeatedly rotate at the requested
key until it becomes the root. The Move-to-Root algorithm enacts this process
at each request. The \emph{crossing bound} for $X$ starting from $T$, denoted
$\CrossingBound(X,T)$, is the crossing cost of Move-to-Root's execution for
this instance. The crossing bound is essentially equivalent to Wilber's second
lower bound on optimal execution cost~\cite{WILBER}. Thus, the crossing bound
never exceeds a fixed multiple of optimal execution cost. (See
Appendix~\ref{app:WilbersLowerBound}.) We shall prove that the crossing bound
is approximately monotone. Our techniques preview those required to show the
same for Splay. We begin with three properties of Move-to-Root. The first
describes the structure of its transition trees.

\begin{theorem}\label{thm:GlobalMoveToRoot}
Executing $\MoveToRoot(T,x)$ transforms the access path to $x$ in $T$ into a
tree whose root $x$ has left and right subtrees that are respectively right and
left spines.
\end{theorem}

\begin{proof}
By induction on the number of rotations involved in the operation. If the
requested key lies at the root then the statement is trivial. Now suppose that
the statement is true for nodes of depth $k$, let $\Depth_T(x)=k+1$ and
$z=\Root(T)$, and assume without loss of generality that $T$ solely comprises
keys on the access path for $x$ in $T$ and that $x<z$. (The other case is
symmetric.) The first $k-1$ of the $k$ successive rotations at $x$ performed
while executing $\MoveToRoot(T,x)$ replace the left subtree $Q$ of $z$ in $T$
with $Q'=\MoveToRoot(Q,x)$. By the inductive hypothesis, the left and right
subtrees of $x$ in $Q'$ comprise inward facing spines of the keys in
$T\setminus\Set{x,z}$ that are respectively less than and greater than $x$. The
left subtree of $x$ remains unchanged after the final rotation, while the right
subtree of $x$ immediately before the final rotation becomes the left subtree
of $z$ immediately afterward, and the right subtree of $x$ after the final
rotation is a left spine of keys greater than $x$. Thus, the hypothesis holds
for nodes at depth $k+1$.
\end{proof}

The second property demonstrates Move-to-Root's executions reflect temporal
patterns in the request sequence. A binary tree is \emph{max-heap ordered} if
each node is assigned a priority from a totally ordered set and every non-root
node's priority is at most that of its parent's. The \emph{standard priorities}
for a request sequence $X=(x_1,\dots,x_m)$ starting from $T$ are the mappings
$p_0,p_1,\dots,p_m$ from keys to priorities such that, for $y\in T$,
$p_0(y)=\tau(y)-|T|-1$ where $\tau(y)$ is the time at which $y$ is requested in
$\Postorder(T)$, and $p_i(y)=p_{i-1}(y)$ if $y\ne x_i$ and $p_i(x_i)=i$ for
$1\le i\le m$.

\begin{theorem}\label{thm:MoveToRootHeap}
Move-to-Root's after-trees are max-heap ordered by the instance's standard
priorities.
\end{theorem}

\begin{proof}
The initial tree is max-heap ordered with respect to the initial priorities:
the root of the initial tree has highest priority, and the same holds
recursively for its subtrees. We can see as follows that Move-to-Root restores
the max-heap order invariant after each-request. Resetting the priority of the
node holding requested key $x$ introduces a single heap order violation at the
edge between $x$ and its parent, if the parent is present. After each rotation
at $x$ that does not result in $x$ becoming the root, only a single edge in the
tree violates the heap order, and that edge is always the one between $x$ and
its parent. When $x$ becomes the root, it has the largest priority, and no
other edges violate the heap order.
\end{proof}

The third property characterizes how Move-to-Root arranges keys in its
after-trees. The left and right \emph{window boundaries} $u$ and $v$ for a
given key $y$ determined by a request sequence $X$ are respectively the largest
key less than or equal to $y$ and the smallest key greater than or equal to $y$
in $X\oplus(-\infty, \infty)$. The \emph{window subtree} $J$ for $y$ determined
by an execution of $X$ starting from $T$ with final tree $R$ is as follows. If
neither $u$ nor $v$ are finite then $J=T$; if $u=v$ then $J=\varnothing$; if
only $u$ is finite, or if both $u$ and $v$ are finite and the final request for
$u$ precedes the final request for $v$ in $X$, then $J$ is the right subtree of
$u$ in $R$; otherwise, $J$ is the left subtree of $v$ in $R$.

\begin{theorem}\label{thm:MoveToRootWindows}
The window subtree for $y$ determined by Move-to-Root's execution of $X$
starting from $T$ comprises the keys in $T$ strictly between the window
boundaries for $y$ determined by $X$.
\end{theorem}

\begin{proof}
By induction on the number of requests. The initial tree and final tree are
identical for executions of the empty sequence, so the statement is true when
there are no requests. Now suppose the statement is true for sequences of up to
$|W|$ requests and that $X=W\oplus(z)$ for some $z\in T$, and assume without
loss of generality that $y\notin W$. Let $u$ and $v$ be the window boundaries
for $y$ determined by $W$, let $I$ be the set of keys in $T$ that are larger
than $u$ and smaller than $v$, let $R$ be the final tree in Move-to-Root's
execution of $W$ starting from $T$, and let $J$ be the window subtree for $y$
determined by this execution. Define $u'$, $v'$, $I'$, $R'$ and $J'$
analogously for $X$.

Consider first when $z\notin I$ so that $I'=I$ and there are no rotations at
any key in $J$ while moving $z$ to the root, and assume without loss of
generality that $J$ is the right subtree of $u$ in $R$. (The case when $J$ is
the left subtree of $v$ is symmetric.) By the inductive hypothesis, $I$ is the
set of keys in $J$. If $z\ne u$ then the right subtree of $u$ in $R'$ is the
same as in $R$ since there are no rotations at $u$ while moving $z$ to the root
of $R$, and $J'$ is the right subtree of $u$ in $R'$ since $v$ is either
infinite or more recently requested in $X$ than $u$. If $z=u$ and $v$ is
infinite then every key greater than $u$ is in its right subtree in $R$ and $u$
is on the right spine of $R$, meaning the right subtree of $u$ in $R'$ is the
same as in $R$ and $J'$ is again the right subtree of $u$ in $R'$. If $z=u$ and
$v$ is finite then $J'$ is the left subtree of $v$ in $R'$ and $v$ is an
ancestor and the successor of $u$ in $R$, meaning the left subtree of $v$ in
$R'$ is $J$ since Move-to-Root is a natural algorithm. In all cases $J'=J$ and
the hypothesis holds for $X$.

Consider now when $z\in I$. If $z=y$ then $J'=I'=\varnothing$ by construction.
Otherwise, assume without loss of generality that $z<y$ so that $u'=z$. (The
case when $v'=z$ is symmetric.) Form $\hat{R}$ by substituting
$\MoveToRoot(J,u')$ for $J$ in $R$ and let $\hat{J}$ be the right subtree of
$u'$ in $\hat{R}$. Since $I'$ and $\hat{J}$ have the same keys and
$R'=\MoveToRoot(\hat{R},u')$, we may apply analysis of when $z\notin I$.
\end{proof}

Our proof of approximate monotonicity examines how individual request removals
affect the crossing bound. In particular, removing the first request in a
sequence subtracts the first key’s crossing depth from the crossing cost of
Move-to-Root's execution of the remaining requests. The remaining requests are
now executed starting from the original tree, rather than the tree resulting
from moving the first requested key to the root. As the remainder of the
altered execution proceeds, its after-trees become progressively similar to
those of Move-to-Root’s execution of the original request sequence. The key to
our argument is bounding the cost incurred by this restoration process.

\begin{theorem}\label{thm:RequestRemoval}
$\CrossingBound(X,T) - \CrossingBound(X,\MoveToRoot(T,y)) \le 4\Level_T(y)$.
\end{theorem}

\begin{proof}
Let $R'$ and $S'$ be the final trees in Move-to-Root's executions of $X$
starting respectively from $T$ and $\MoveToRoot(T,y)$, let $J'$ and $K'$ be the
window subtrees for $y$ determined by these executions, and set $k'$ to be the
crossing depth of $y$ in $J'$ if $J'$ is nonempty and zero otherwise. We show
by induction on the number of requests that $\CrossingBound(X,T) -
\CrossingBound(X,\MoveToRoot(T,y)) \le 4(\Level_T(y)-k')$. The statement is
true by construction for the empty sequence, so consider when $X=Y\oplus(z)$
for some $z\in T$. Define $R$, $S$, $J$, $K$ and $k$ analogously for $Y$, let
$I$ be the set of keys in $T$ contained in the symmetric order interval
strictly between the window boundaries for $y$ determined by $Y$, and suppose
$\CrossingBound(Y,T) - \CrossingBound(Y,\MoveToRoot(T,y)) \le 4(\Level_T(y)-k)$.

Since $\CrossingBound(X,T) - \CrossingBound(X,\MoveToRoot(T,y)) =
\CrossingBound(Y,T) - \CrossingBound(Y,\MoveToRoot(T,y)) +
\Level_R(z)-\Level_S(z)$ it suffices to show that $\Level_R(z)-\Level_S(z)\le
4(k-k')$. Thus, we need to characterize the structure of the final trees. The
tree $\MoveToRoot(T,y)$ is max-heap ordered with respect to a priority function
that is identical to the standard priority function for starting tree $T$
except at $y$, whose priority in $\MoveToRoot(T,y)$ we set by convention to
zero. By Theorem~\ref{thm:MoveToRootHeap}, the after-trees of Move-to-Root's
executions of $X$ starting from $T$ and $\MoveToRoot(T,y)$ are max-heap ordered
by priority functions defined recursively in the standard way starting
respectively from the priority functions for $T$ and $\MoveToRoot(T,y)$.
Requests subsequent to $y$ have the same after-trees in both executions, so we
may assume without loss of generality that $y\notin Y$ and $k>0$.

By Theorem~\ref{thm:MoveToRootWindows}, when $Y\ne\varnothing$ the keys in
$T\setminus I$ comprise a rooted hull in $R$ and $S$. (If $Y=\varnothing$ then
$T\setminus I$ is empty.) These keys have the same priorities in both $R$ and
$S$, meaning the two rooted hulls are identical. The window subtree $J$ has the
same keys as $K$, none of which are requested in $Y$, meaning these keys have
their initial priorities in $R$ and $S$. The only key with differing priority
in $J$ and $K$ is $y$, which has maximal priority among keys in $K$. Therefore,
$K=\MoveToRoot(J,y)$.

If $z\notin I$ then $k'=k$ and $\Level_R(z)=\Level_S(z)$, so we may assume
$z\in I$ without loss of generality. If $J=R$ set $J^+=J$, otherwise set $J^+$
to be the subgraph in $R$ comprising the union of keys in $J$ with the window
boundary for $y$ determined by $Y$ of which the root of $J$ is a child in $R$.
Define $K^+$ analogously for $K$ in $S$. The access path to $\Root(J^+)$ in $R$
is identical to the access path for $\Root(K^+)$ in $S$ whenever $Y$ is
nonempty, and the access paths to this key in $R$ and $S$ are identical. Thus,
$\Level_R(\Root(J^+))=\Level_S(\Root(K^+))$ while $\Level_R(z) =
\Level_R(\Root(J^+)) + \Level_{J^+}(z) - 2$ and $\Level_S(z) =
\Level_S(\Root(K^+)) + \Level_{K^+}(z) - 2$, meaning
$\Level_R(z)-\Level_S(z)=\Level_{J^+}(z)-\Level_{K^+}(z)$.

Let $P$ be the be the smallest rooted hull in $J$ containing the neighborhood
of $z$ in $J$. Since Move-to-Root is a natural algorithm, the subtrees hanging
from $P$ are unaffected by executing $\MoveToRoot(J,y)$, and so these subtrees
are identically arranged in $K$. Therefore, we may assume with no loss of
generality that either $z\in P$ or that the parent of $z$ in $R$ is defined and
lies in $P\setminus\Set{y}$.

Let $w_1,\dots,w_{k-1}$ be the first $k-1$ crossing nodes for $y$ in $J$ in
increasing order of depth and let $w_0=y$. Since $J=K$ and $R=S$ if $y$ is the
root of $J$, we may may assume without loss of generality that $y$ has a parent
in $J$. Thus, let $w_k$ be the child of $y$ in the same direction as $y$ with
respect to its parent in $J$ and let $w_{k+1}$ be the child of $y$ in the
opposite direction. Define $j=\Level_J(w_c)$ where $c$ is the largest integer
for which $w_c$ is an ancestor of $z$ in $J$ if $z\ne y$ and otherwise $c=0$.
Let $D$ be the indicator for $Y\ne\varnothing$, let $A$ be the indicator for
$z\notin P$, let $B$ be the indicator for $z\ne w_c$, let $G$ be the indicator
for $\Level_J(y)<\Level_{J^+}(y)$ and let $F$ be the indicator for
$\Level_K(z)<\Level_{K^+}(z)$. (An \emph{indicator}'s value is one when its
condition is true and zero otherwise.) By Theorem~\ref{thm:GlobalMoveToRoot}
and case analysis,
\begin{align*}
 \Level_{J^+}(z) = j +
  \begin{cases}
    G                      & c = 0 \\
    H                      & c = 1 \\
    B(1+A)+G               & \text{otherwise}
  \end{cases}
 && \text{and} &&
 \Level_{K^+}(z) =
  \begin{cases}
    1 + D                  & c = 0 \\
    2 + F + B(1+A)         & 1 \le c \le 2 \\
    3 + F + A              & \text{otherwise,}
  \end{cases}
\end{align*}
where $H=A(1-B)(1+D(1-G))+B(1+A+G)+(1-A)(1-B)D$. (See
Figure~\ref{fig:FunnelSplit}.)

\begin{figure}
\centering
\includegraphics[width=0.8\columnwidth]{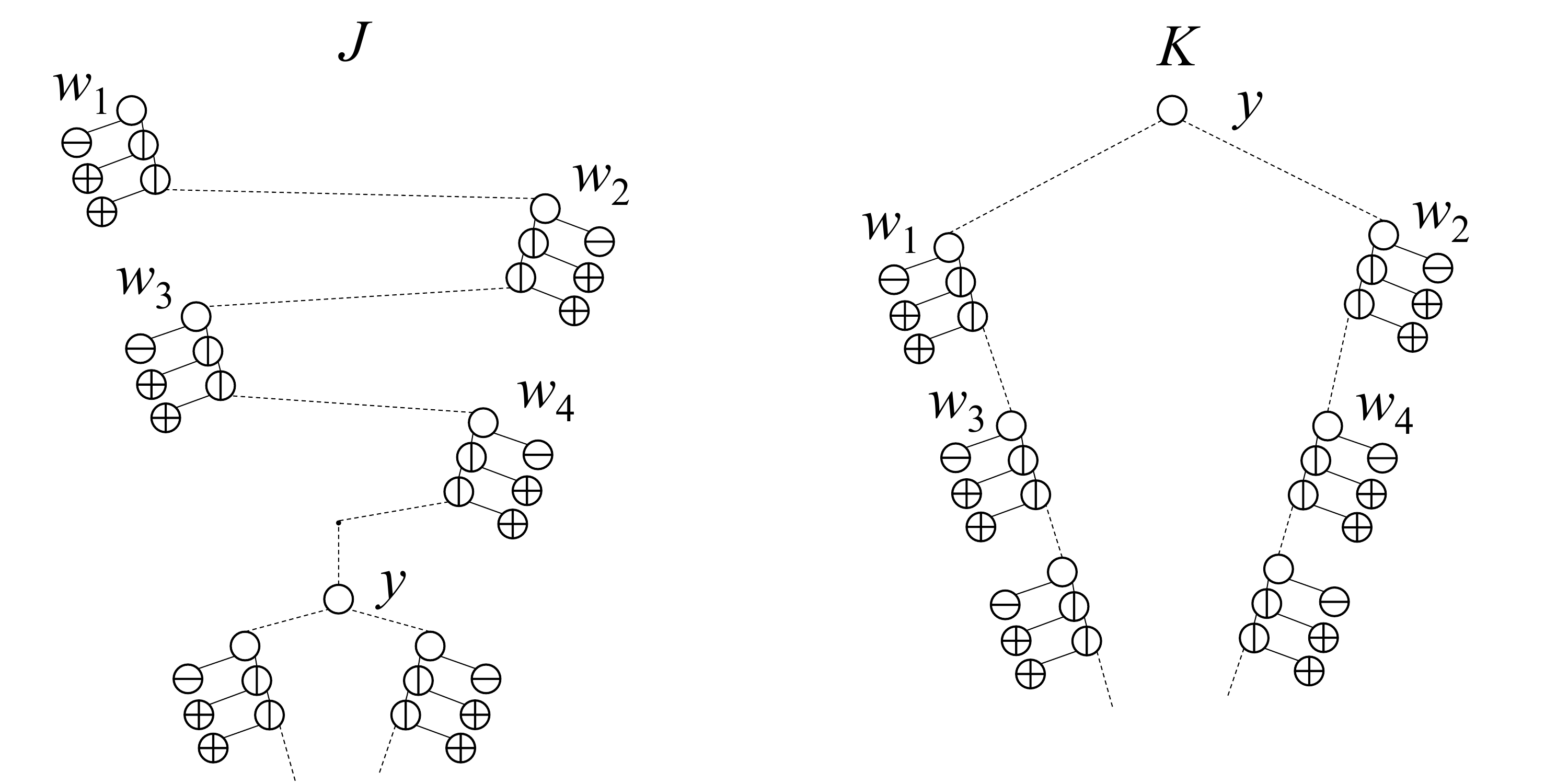}
\caption{Schematic of the window subtrees $J$ and $K$. If $z$ is a horizontally striped node then $A=1$. If $z$ is a vertically striped node then $B=1$, and if $z$ is a cross-hatched node then $AB=1$.}
\label{fig:FunnelSplit}
\end{figure}

Note that if $c=1$ then $F=G$, and if $c=2$ then $F=D(1-G)$, and if $G=1$ or
$F=1$ then $D=1$. Combining these facts with some case analysis reveals that if
$1\le j\le 2$ then $\Level_{J^+}(z)-\Level_{K^+}(z)=0$, and otherwise
$\Level_{J^+}(z)-\Level_{K^+}(z)\le j$. If $c=0$ then $k'=0$ and $j=k$.
Otherwise, we apply Theorem~\ref{thm:GlobalMoveToRoot} to deduce the structure
of the access path $P'$ for $y$ in $J'$. If $c=1$ then $P'$ is a subgraph of
$P$ and $k'\le k$. If $z$ is in the subtree rooted at $w_k$ then $P'$ is a
right spine ending at $y$ and if $z$ is in the subtree rooted at $w_{k+1}$ then
$P'$ is a left spine ending at $y$. Therefore when $z$ is a strict descendant
of $y$, if $k=2$ and $c=3$ then $k'=1$ since $w_{k+1}$ and the parent of $y$ in
$P$ are on the same side of $y$ in symmetric order, and otherwise $k'\le 2$ and
$j\le k+1$. Otherwise, $2\le c\le k-1$, in which case if $z\in P$ then set $q$
to be the child of $z$ child in $P$ and otherwise set $q$ to be the parent of
$z$ in $P$. Let $U$ be a right spine comprising the strict ancestors of $q$ in
$P$ that are less than $y$ and let $V$ be a left spine of the strict ancestors
of $q$ in $P$ that are greater than $y$, and let $M$ be the path from $q$ to
$y$ in $P$. If $z<y$ then $P'$ results from attaching $M$ as the left subtree
of the node with smallest key in $V$, and if $z>y$ then $y$ then $P'$ results
from attaching $M$ as the right subtree of the node with largest key in $U$.
Thus, $k'\le k-j+2$. In all cases, $\Level_{J^+}(z)-\Level_{K^+}(z)$ is at most
$4(k-k')$.
\end{proof}

\begin{theorem}\label{thm:MonotoneCrossingBound}
The crossing bound is approximately monotone.
\end{theorem}

\begin{proof}
Let $T_1,\ldots,T_m$ be the after-trees of Move-to-Root's execution of
$X=(x_1,\dots,x_m)$ starting from $T$, and let $m\ge e_1>e_2>\cdots>e_p\ge 1$
be a sequence of request times. Set $X_0=X$ and for $1\le i\le p$ form $X_i$ by
removing request $e_i$ from $X_{i-1}$. We induct on $p$ to show that
$\CrossingBound(X_p,T)\le \CrossingBound(X,T)+3\sum_{i=1}^p
\Level_{T_{e_i-1}}(x_{e_i})$ where $T_0=T$, which suffices to establish that
the crossing bound has subsequence overhead at most four. The statement is
trivial when $p=0$. Now suppose $\CrossingBound(X_{p-1},T) \le
\CrossingBound(X,T) + 3\sum_{i=1}^{p-1}\Level_{T_{e_i-1}}(x_{e_i})$. Let
$y=x_{e_p}$ and let $W$ and $Z$ respectively be the first $e_p-1$ and final
$m-e_p-(p-1)$ requests in $X_{p-1}$, so that $X_{p-1}=W\oplus(y)\oplus Z$ and
$X_p=W\oplus Z$. Let $S=T_{e_p-1}$, so that $\CrossingBound(X_{p-1},T) =
\CrossingBound(W,T) + \CrossingBound((y)\oplus Z,S)$ and $\CrossingBound(X_p,T)
= \CrossingBound(W,T) + \CrossingBound(Z,S)$. Note that
$\CrossingBound((y)\oplus Z,S) = \Level_S(y) +
\CrossingBound(Z,\MoveToRoot(S,y))$. Thus, $\CrossingBound(X_p,T) -
\CrossingBound(X_{p-1},T) = \CrossingBound(Z,S) -
\CrossingBound(Z,\MoveToRoot(S,y)) - \Level_S(y)$, which by
Theorem~\ref{thm:RequestRemoval} is at most $3\Level_S(y)$. Therefore,
$\CrossingBound(X_p,T) \le \CrossingBound(X_{p-1},T) +
3\Level_{T_{e_p-1}}(x_{e_p})$, and the hypothesis holds for $p$ request
removals.
\end{proof}

Our presentation of the crossing bound is based on Iacono's
work~\cite{IN_PURSUIT,KEY_INDEPENDENCE}. Move-to-Root, introduced by Allen and
Munro~\cite{MOVE_TO_ROOT}, is the earliest example of a self-adjusting binary
search tree algorithm. The crossing bound is not strictly monotone. For
example, $\CrossingBound(Y,T) > \CrossingBound(X,T)$ when $X=(4,5,3)$,
$Y=(5,3)$ and $\Postorder(T)=(3,2,5,6,4,7,1)$. We had not realized this when
writing our SODA paper, whose treatment of the crossing bound contains several
mistakes~\cite{A_NEW_PATH}.

\section{The Way Forward}\label{sec:TheWayForward}

Our numerical experiments indicate that Splay's cost never exceeds four times
the sum of an instance's crossing bound and initial tree size, which would
imply dynamic optimality. Moreover, the crossing costs of Splay and
Move-to-Root are so tightly coupled that the difference between them may well
be at most linear in initial tree size. Additionally, the keys in the crossing
nodes of these algorithms' executions are quite similar, albeit sometimes
offset from each other in symmetric order by a small amount. We have tried to
prove these statements, to no avail. We believe our failures are not
incidental, and that there are structural obstacles in the way of establishing
dynamic optimality in this manner. The difficulty arises from temporal spread.
Typically, about half of the keys in Move-to-Root's crossing nodes for a given
request appear on the access path for the corresponding request in Splay's
execution. A smaller fraction of these keys appear on the splay path for the
next request, and the remaining keys are scattered across subsequent splay
paths. The precise extent of this spreading is varied and depends on the
particular request sequence.

Lucas remarked that optimal cost does not seem amenable to inductive
analysis~\cite{CANONICAL_FORMS}. The observed temporal mixing is a
manifestation of this problem, since it means that showing Splay's cost obeys
the crossing bound requires accounting for many of its preceding transition
trees at each request. Fortunately, dynamic optimality's equivalence to
approximate monotonicity provides a means of shattering this barrier. Our
conviction is:

\begin{conjecture}\label{conj:MonotoneSplayCrossings}
Splay's crossing cost is approximately monotone.
\end{conjecture}

Our proof of the crossing bound's monotonicity is a natural starting point for
tackling Conjecture~\ref{conj:MonotoneSplayCrossings}. However, it requires a
crucial modification. Our analysis of Move-to-Root establishes a worst-case
bound on how its crossing cost increases after removing a request. By contrast,
the increase in Splay's crossing cost is not bound by any fixed multiple of the
removed key's crossing depth. For example, if $T$ results from splaying the
largest key in a right spine comprising the integers $\Set{1,\dots,2n}$,
$Y=(2,4,\dots,2n-2)$ and $X=(1)\oplus Y$, then the crossing costs of Splay's
executions of $X$ and $Y$ when starting from $T$ are respectively $3n-4$ and
$5n-5$. Consequently, we must examine how request removals affect Splay's
crossing cost in aggregate, which is equivalent to reversing the order in which
the proof of Theorem~\ref{thm:MonotoneCrossingBound} inducts on request
removals. This style of induction entails comparing executions of a request
sequence starting from progressively divergent trees. The increased complexity
of the required new approach is another manifestation of the barriers to
standard inductive analysis of optimal algorithms. The advantage of attacking
this manifestation of the problem is that
Theorem~\ref{thm:MonotoneCrossingBound} assures its achievability.

Adapting such a proof from Move-to-Root to Splay will almost certainly require
a \emph{potential function} in order to smooth out the effects of occasional
requests whose removal produces a high increase in Splay's crossing cost.
Potential functions are tools for analyzing algorithms that have individual
operations with high cost, but for which the cost per operation, amortized over
all operations in a sequence, is low~\cite{AMORTIZED_COMPLEXITY}. Each possible
configuration of the data structure (e.g.\ the tree) is assigned a numerical
value, called its \emph{potential}. The cost of an operation is redefined to
depend on both the original cost (e.g.\ the length of the Splay path), and on
how the potential changes due to the operation's effect on the data structure.
If carefully constructed, the sum of the redefined costs over a sequence of
operations will be an upper bound on the sum of the actual costs, yet no
individual operation's redefined cost will ever be very large. We need a
potential that captures how Splay's executions diverge from Move-to-Root's.

\begin{figure}
\centering
\includegraphics[width=0.8\columnwidth]{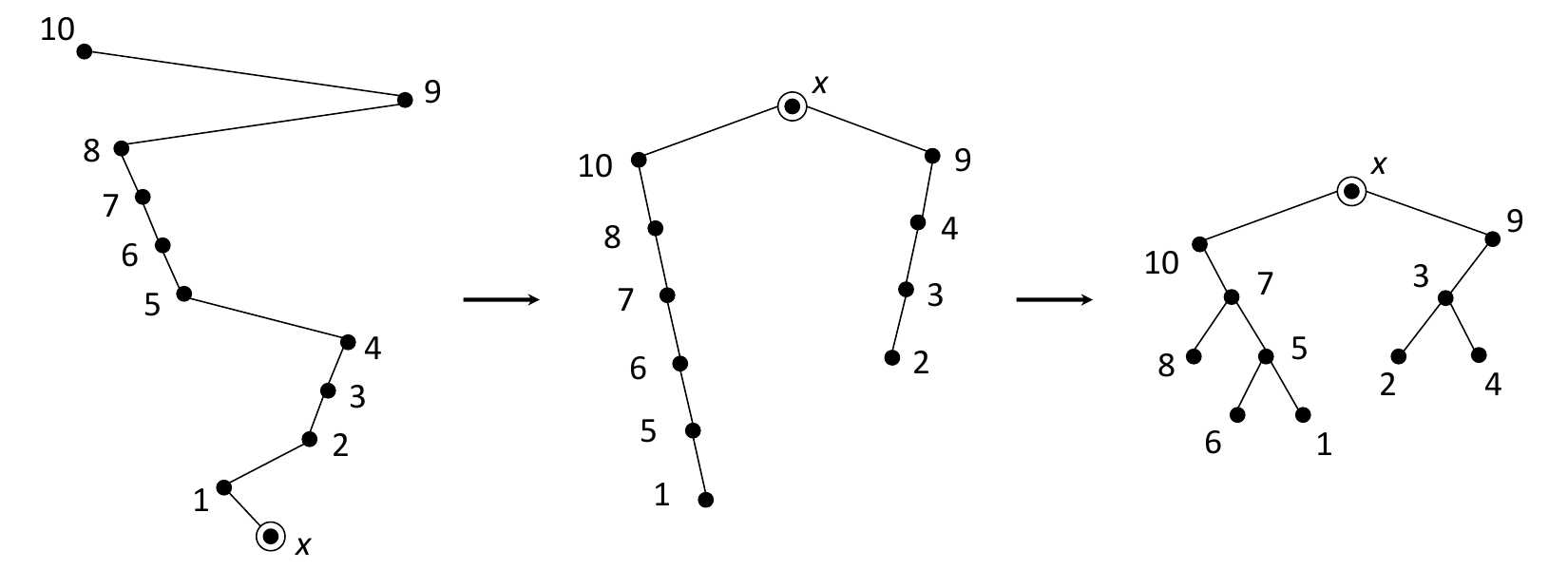}
\caption{The transformation from the left to the middle illustrates Move-to-Root. The transformation from the left to the right illustrates Splay. (Illustration from~\cite{GLOBAL_VIEW}.)}
\label{fig:GlobalSplay}
\end{figure}

Move-to-Root is a both progenitor and a sub-step of splaying. Move-to-Root
splits the access path into a pair of spines. One can view Splay as comprising
two phases: the first executes move-to-root, the second performs extra
rotations, corresponding to the zig-zigs. (See Figure~\ref{fig:GlobalSplay}.)
The extra rotations ensure that a splay operation decreases the depth of every
node by about half of the number of its ancestors that were on the access path
for the requested key~\cite{EXPLANATION_OF_SPLAYING}.

\begin{theorem}[{\cite[Proposition~17]{GLOBAL_VIEW}}]\label{thm:GlobalSplay}
Executing $\Splay(T,x)$ is equivalent to starting from the tree
$\MoveToRoot(T,x)$ and rotating at every key held by a child $y$ on the access
path for $x$ in $T$ whose parent in $T$ is on the same side of $x$ in symmetric
order for which $\Depth_T(x)-\Depth_T(y)$ is odd.
\end{theorem}

\begin{proof}
By induction on the number of splay steps involved. The theorem is trivially
satisfied when splaying at the root. Now suppose the statement is true for
splay operations comprising $k-1$ splay steps and let $x$ be a node in $T$
whose splaying involves $k$ steps. Let $Q$ be the subtree rooted at the
ancestor of $x$ in $T$ whose depth is $\Depth_T(x)-2(k-1)$. Since the first
$k-1$ splay steps each decrease the depth of $x$ by two,
$\Splay(T,x)=\Splay(S,x)$ where $S=\Splay(Q,x)$. Denote by $\Splay'$ the
procedure described in the theorem. Let $z=\Root(S)$, let $y$ be the left child
of $z$ in $S$ and assume without loss of generality that $x<z$. (The other case
is symmetric.) If $x=y$ then $\Splay'(S,x)$ only enacts a single rotation at
$x$, which is equivalent to the zig enacted by $\Splay$. If $x>y$ then $y$ and
$z$ are on opposite sides of $x$ in symmetric order and $\Splay'(S,x)$ rotates
twice at $x$, making it identical to the zig-zag step performed by $\Splay$.
Finally, if $x<y$ then $\Splay'(S,x)$ rotates twice at $x$ and then once at
$y$, which is equivalent the rotation at $y$ followed by $x$ in the zig-zig
performed by $\Splay$. In all three cases, $\Splay'(S,x)=\Splay(S,x)$. Form
$S'$ by replacing $Q$ with $\Splay'(Q,x)$ in $T$. The edges rotated by
$\Splay'$ subsequent to the $\MoveToRoot$ operation are disjoint, and
$\Depth_{S'}(x)-\Depth_{S'}(w)$ and $\Depth_T(x)-\Depth_T(w)$ have the same
parity for every ancestor $w$ of $x$ in $S'$. Therefore,
$\Splay'(T,x)=\Splay'(S',x)$. By the inductive hypothesis $S'=S$, thus
$\Splay'(T,x)=\Splay(T,x)$.
\end{proof}

Each zig-zig can create a violation of the max-heap ordering with respect to
the standard priorities of an instance. As the executions of both Splay and
Move-to-Root proceed, these zig-zigs will sometimes create further heap order
violations. At other times, splay steps will remove some of the heap order
violations. The correct potential for analyzing Splay's crossing cost should in
some way bound the rate at which splay operations generate heap order
violations with respect to the standard priorities. We speculate on two
possible forms. The first simply counts the number of edges in the tree being
splayed that violate the heap-order condition with respect to most recent
access time. This potential may be too ``coarse,'' in that it fails to capture
heap order violations between nodes not immediately connected by an edge. If
so, the likely way to address this shortcoming is weighting each node by some
function of the difference between its crossing depth in the splayed tree and
in the max-heap order maintained by Move-to-Root. While honing the details of
the potential's construction falls outside this work's scope, we can infer
something important up front.

A potential function's design is closely tied to the extent to which its value
can increase or decrease. By Theorem~\ref{thm:SplayIsRepeatable}, if Splay is
dynamically optimal then its startup overhead is no more than linear in initial
tree size. Hence, any potential for proving
Conjecture~\ref{conj:MonotoneSplayCrossings} should also have maximum value at
most linear in the size of the starting tree. This considerably narrows the
design space that we might otherwise need to explore.

To prove optimality we must also address Splay's bookkeeping cost. Here again,
we can glean insight from Splay's progenitor. Move-to-Root is not dynamically
optimal. For example, if $T$ is a left spine with keys $\Set{1,\dots,n}$,
$X=(n,n-1,\dots,2,1,2,\dots,n-1,n)$ and $Y=(1,2,\dots,n)$ then the cost of
executing Move-to-Root on the subsequence $Y$ is proportional to $n$ times the
cost of its execution on the super-sequence $X$. Because its crossing cost
lower bounds a fixed multiple of optimal cost, Move-to-Root's non-optimality
arises from its bookkeeping cost. Splay tweaks Move-to-Root by breaking apart
bookkeeping edges via zig-zig steps. Thus, Splay seems to be precisely the
modification needed to make Move-to-Root optimal. We believe:

\begin{conjecture}\label{conj:SplayBookkeeping}
Splay's bookkeeping cost is at most a fixed multiple of the sum of its crossing
cost and initial tree size.
\end{conjecture}

Heuristically, splaying in a tree whose access paths comprise mostly
bookkeeping nodes increases the average crossing depth of nodes in the tree,
and the opposite phenomenon occurs in trees with many nodes of high crossing
depth. Precisely tracking this exchange as Splay's execution progresses quickly
becomes unmanageable, indicating the need for an additional potential function
that acts as a proxy for the number of bookkeeping nodes in the tree being
splayed. It seems likely that a tree entirely comprising a spine should
maximize this potential, and that a perfectly balanced binary search tree
should minimize it. Conjectures~\ref{conj:MonotoneSplayCrossings}
and~\ref{conj:SplayBookkeeping} together imply dynamic optimality.

\appendix
\section{Rotational Execution}\label{app:RotationalExecution}

Sleator and Tarjan~\cite{SPLAY_JOURNAL}, in their formulation of the dynamic
optimality conjecture, use a rotation-based definition of a binary search tree
execution. Given an initial tree and a request sequence, a \emph{rotational}
execution fulfills one request at a time, by performing a binary search for the
requested key in the current tree, at a cost equal to the number of nodes on
the access path. In addition, the execution can include any number of rotations
before each request, at a cost of one per rotation. Formally, a rotational
execution $R$ of $X=(x_1,\ldots,x_m)$ starting from $T$ comprises a sequence of
trees $T_0,T_1,\dots,T_r$ and search times $0\le \tau_1\le \cdots \le \tau_m =
r$ where $T_0=T$ and $T_t$ results from rotating at some key in $T_{t-1}$ for
$1\le t\le r$. The cost to execute $R$ is $r+\sum_{i=1}^m
\Depth_{T_{\tau_i}}(x_i)$. We denote the cost of an optimal rotational
execution for this instance by $\OPTrot(X,T)$. We shall prove that any
transition tree execution can be simulated by a rotational execution of at most
twice the cost, and vice-versa. Thus the two cost models are the same to within
a factor of two.

\begin{theorem}\label{thm:ModelEquivalence}
$1/2\le\OPT(X,T)/\OPTrot(X,T)\le 2$.
\end{theorem}

\begin{proof}
First we observe that any transition tree execution can be simulated by a
rotational execution at a cost of a factor of at most two. A $k$-node binary
search tree with $k$ keys can be transformed into any other binary search tree
of the same set of keys by doing at most $2k-2$
rotations~\cite[Theorem~2.1]{ORIGINAL_ROTATION_DISTANCE}. Hence each successive
after-tree in the transition tree model can be produced from the previous one
by doing at most $2k-2$ rotations, where $k$ is the number of nodes in the
rooted hull (and in the corresponding transition tree). Searching for the
desired key after doing these rotations costs one. Hence if a transition tree
execution fulfills a request with a transition tree of size $k$, then a
rotational executional execution can fulfill this request with cost at most
$2k-1<2k$.

Simulating a rotational execution by a transition tree execution is more
complicated, because the former allows rotations to be done anywhere in the
tree, not just in a rooted hull. The first step toward handling this is to view
edges as retaining their identity throughout a rotational execution. Consider a
rotation of a left child $x$ whose parent is $y$; let $u$, $v$, and $w$ be the
left child of $x$, the right child of $x$, and the right child of $y$,
respectively. Let $e$, $a$ and $b$ be the edges connecting $x$, $v$ and $y$
with their parents, respectively. (If $v=\Null$ then $a=\Null$ and if $y$ is
the root then $b=\Null$.) Rotation at $e$ swaps the ends of $e$ and converts it
from a left edge to a right edge, converts $a$ from a right edge to a left edge
and changes its top end from $x$ to $y$, and changes the bottom end of $b$ to
$x$. (Rotation at $e$ does not change $\Null$ edges.) The rotation affects no
other edges, and it preserves the set of keys in the subtree rooted at any node
other than $x$ and $y$, and in particular those rooted at $u$, $v$, and $w$.
Right rotations behave symmetrically.

The second step is to modify the rotational execution so that whenever a key is
searched for it is at the root of the tree. Before a search occurs, we first
rotate on each edge of the access path, bottom-up, which moves the key to be
searched for to the root; then we perform the search; then we do the inverse
rotations in the opposite order, restoring the original access path. Fulfilling
the request in this way costs $2k-1$ if the original access path has $k$ nodes.
Thus we increase the overall cost by at most a factor of two.

Finally, assume that a rotational execution moves each requested key to the
root before searching for it. We simulate this rotational execution with a
transition tree execution while at the same time postponing some rotations.
Proceeding in the same order as keys are requested, we modify the subsequence
of rotations before the first search, and each subsequence of rotations between
successive searches, as follows. Let $S$ be such a subsequence, let $T$ be the
tree in which these rotations begin, and let $U$ be the subgraph of $T$
comprising the edges in $S$. We partition $S$ into pair of subsequences $A$ and
$B$. The subsequence $A$ comprises rotations in $S$ at edges in the same
connected component of $U$ as the root of $T$. (If no such rotations are
present then $A$ is empty.) The subsequence $B$ is the complementary
subsequence to $A$ in $S$. We replace $S$ with $A\oplus B$ unless $S$ is the
subsequence of rotations for the final request, in which case we replace $S$
with $A$ and drop the remaining rotations. Then, we move the search time for
the request to occur immediately after the final rotation in $A$.

If $A$ is nonempty then its edges comprise a rooted hull in $T$, and the
rotations in $A$ transform this rooted hull into a tree on the same set of keys
whose root contains the requested key. The transformed tree is the transition
tree corresponding to the request in the transition tree execution. (If $A$ is
empty then the transition tree comprises solely the root of $T$.) If the rooted
hull (and the transition tree) contain $k$ nodes, the number of rotations is at
least $k-1$, making the cost of these rotations plus the cost of the search at
least $k$ in the rotational execution. The size of the corresponding transition
tree is $k$. We conclude that it is possible to simulate a rotational execution
whose searches occur at the root with a transition tree execution of the same
cost, and at most twice the cost for a general rotational execution. Creating
simulations for optimal executions of each type establishes the result.
\end{proof}

Wilber was the first to restrict rotational executions to search only at the
root~\cite{WILBER}. The procedure for partitioning rotations is implicit in
Lucas' work~\cite{CANONICAL_FORMS}. Our description is based on Koumoutsos'
remarks~\cite{ADVICE}. Harmon was the first to describe binary search tree
executions using transition trees~\cite{HARMON_THESIS}.

\section{Wilber's Lower Bound}\label{app:WilbersLowerBound}

We show that the crossing bound is at most a fixed multiple of optimum
transition tree execution cost. Our proof proceeds in two main steps. First, we
express a scoring procedure defined by Wilber in terms of the crossing bound.
Then we use Wilber's proof that this procedure lower bounds optimum rotational
cost as a black box in our analysis to obtain the desired result. (Wilber's
proof is quite intricate, and we do not attempt to summarize it.) Unlike the
crossing bound, Wilber's scoring procedure depends only on the request
sequence. Accounting for initial trees requires care.

Formally, \emph{Wilber's bound} for request sequence $X=(x_1,\dots,x_m)$,
denoted $\Wilber(X)$, is $m+\sum_{i=1}^m \kappa(X,i)$, where the \emph{score}
$\kappa(X,i)$ for each request $1\le i\le m$ is as follows. If $i=1$ then the
score is zero. Otherwise, let $c_1=i-1$ and let $w_1=x_{i-1}$. If $w_1<x_i$ set
$v_0=\infty$, otherwise set $v_0=-\infty$. Initialize $l=1$ and repeat the
following process for as long as $w_l\ne x_i$ and there are keys requested
prior to time $c_l$ lying between $x_i$ (inclusive) and $v_{l-1}$ (exclusive)
in symmetric order. Set $c_{l+1}$ to the latest request time preceding $c_l$
for a key lying between $x_i$ (inclusive) and $v_{l-1}$ (exclusive) in
symmetric order. Set $w_{l+1}$ to the key requested at $c_{l+1}$. Set $v_l$ to
the key closest in symmetric order to $x_i$ (exclusive) on the same side of
$x_i$ in symmetric order as $w_l$ that is requested after $c_{l+1}$ and no
later than $c_l$. Finally, increment $l$ by one. The score is one less than the
terminal value of $l$. We respectively refer to $w_1,\ldots,w_l$ and
$v_0,\ldots,v_{l-1}$ as the crossing keys and inside keys for the request.
Wilber's bound is nearly the same as the crossing bound for $X$ starting from
the \emph{default tree} $\BST(X)$ comprising the keys in $X$ min-heap ordered
by their first request times.

\begin{theorem}\label{thm:CrossingNodeEquivalence}
$\Wilber(X) = \CrossingBound(X,\BST(X)) - |\BST(X)| + 1$ whenever $X \ne
\varnothing$.
\end{theorem}

\begin{proof}
By induction on the number and crossing depths of requests. Since the first
request's score is zero, Wilber's bound is one for the singleton request
sequence. Meanwhile, the first requested key lies at the root of the default
tree for the request sequence and the root has crossing depth one. Thus, the
formula holds for sequences containing a single request. Now suppose the
theorem is true for all request sequences of length up to $m-1$, let $Y$ be a
nonempty sequence of $m-1$ requests, let $X=Y\oplus(x)$, let $T$ be the final
after-tree in Move-to-Root's execution of $Y$ starting from $\BST(X)$, and set
$\delta$ to be one if $x\notin Y$ and zero otherwise. We show that the first
$\Level_T(x)-\delta$ crossing nodes for $x$ in $T$, ordered increasing by
depth, contain the crossing keys for request $m$, and that the respective
parents of these nodes contain the inside keys for the request. (If the zeroth
inside key is $\infty$ we treat $T$ as the left subtree of this key, and
otherwise as the right subtree of $-\infty$.)

The last key requested in $Y$ is the first crossing key for request $m$ in $X$.
Meanwhile, by Theorem~\ref{thm:MoveToRootHeap}, the keys in $Y$ comprise a
rooted hull in $T$ max-heap ordered by their last request times in $Y$. In
particular, the root of $T$, which is the first crossing node for $x$ in $T$,
contains the first crossing key. Now suppose that the first $i$ crossing nodes
for $x$ in $T$ contain the first $i$ crossing keys for request $m$ in $X$ for
some $1\le i<\Level_T(x)$, and that the parents of these nodes contain the
first $i$ inside keys. Let $w$ and $w'$ respectively be the deepest among the
first $i$ and $i+1$ crossing nodes for $x$ in $T$, let $v$ and $v'$ be the
respective parents of these nodes, and assume without loss of generality that
$x<w$. (The other case is symmetric.)

First consider when $w'\in Y$. Since $w'$ is a descendant of $w$ in $T$, the
former's final request time in $Y$ precedes the latter's. Because $w'$ is in
the right subtree of $v$ and either $w'=x$ or $w'$ contains $x$ in its right
subtree, $w'$ is greater than $v$ and at most $x$. Every key in this interval
is a descendant of $w'$, making $w'$ the last among them requested in $Y$.
Applying the inductive hypothesis that $w$ and $v$ respectively contain
crossing key $i$ and inside key $i-1$ for request $m$ in $X$ establishes that
$w'$ contains crossing key $i+1$. Since $v'$ is both the parent of $w'$ and the
deepest node on the left spine of the subtree of $T$ rooted at $w$ which
contains $x$ in its left subtree, it has the smallest key greater than $x$
whose final request comes after the last request for $w'$ and no later than the
last request for $w$ in $Y$. Thus, $v'$ is inside key $i$ for request $m$ in
$X$. Furthermore, if $i+1=\Level_T(x)$ then $w'=x$ and there are no further
crossing keys for request $m$ in $X$.

Otherwise, if $w'\notin Y$ then $w'=x$, $i=\Level_T(x)-1$, and the subtree
rooted at $x$ in $T$ contains every key that is greater than $v$ and at most
$x$. Since Move-to-Root is a natural algorithm and $x$ has no children in
$\BST(X)$, the absense of $x$ in $Y$ ensures that $x$ has no children in $T$,
making $x$ the only key in this interval. Thus, there are only $\Level_T(x)-1$
crossing keys for request $m$ in $X$ when $\delta=1$.

By the inductive hypothesis on request sequences of length $m-1$, $\Wilber(Y) =
\CrossingBound(Y,\BST(Y)) - |\BST(Y)| + 1$, and by the above arguments
$\kappa(X,m) = \Level_T(x) - 1 - \delta$. Since $\Wilber(X) = \Wilber(Y) +
\kappa(X,m) + 1$ and $\CrossingBound(X,\BST(X)) - |\BST(X)| =
\CrossingBound(Y,\BST(Y))-|\BST(Y)| + \Level_T(x)-\delta$, the formula holds
for request sequences of length $m$.
\end{proof}

\begin{theorem}\label{thm:WilbersBound}
$\CrossingBound(X,T) \le 44\OPT(X,T)$.
\end{theorem}

\begin{proof}
Let $P=\Postorder(T)$ and $T'=\BST(P\oplus X)$ and note that $T'=\BST(P)$ since
$P$ contains every key in $T$. Let $A$ and $B$ respectively be optimal
executions for $P$ and $P\oplus X$ starting from $T$, let $U$ be the sequence
comprising the first $|T|-1$ transition trees of $A$, let $V$ be the sequence
comprising the final $|X|$ transition trees in $B$, and let $T''$ be the
after-tree for request $|T|$ in $B$. The transition tree sequence
$U\oplus(T'')\oplus V$ is an execution for $P\oplus X$ starting from $T$ with
cost at most $\OPT(P,T)+|T|+\OPT(X,T'')$. By~\cite[Theorem~4]{SPLAY_PATTERNS},
$\Cost(P,T')\le 7|T|$. Splay's cost and initial tree size respectively upper
bound and lower bound optimum cost, meaning $\OPT(P,T) \le \OPT(P,T')+|T|\le
8|T|$ and $\OPT(X,T'') \le \OPT(X,T) +|T|$. Combining these inequalities
establishes $\OPT(P\oplus X,T) \le \OPT(X,T)+10|T|\le 11\OPT(X,T)$. By
Theorem~\ref{thm:MoveToRootHeap}, $T$ is the final tree in Move-to-Root's
execution of $P$ starting from $T'$. Hence, $\CrossingBound(P\oplus X,T') =
\CrossingBound(P,T') + \CrossingBound(X,T)$. By
Theorem~\ref{thm:CrossingNodeEquivalence}, $\Wilber(P\oplus X) =
\CrossingBound(P\oplus X, T') - |T'|+1$ and $\Wilber(P) =
\CrossingBound(P,T')-|T'|+1$. Therefore, $\CrossingBound(X,T) =
\CrossingBound(P\oplus X,T') - \CrossingBound(P,T') = \Wilber(P\oplus X) -
\Wilber(P) \le \Wilber(P\oplus X)$. Finally, by~\cite[Theorem~7]{WILBER} and
Theorem~\ref{thm:ModelEquivalence}, $\Wilber(P\oplus X) \le 2\OPTrot(P\oplus
X,T) \le 4\OPT(P\oplus X,T) \le 44\OPT(X,T)$.
\end{proof}

\begin{acks}
We thank Lu\'is Russo for suggesting improvements to Figure~\ref{fig:HamCycle},
Kurt Mehlhorn for simplifying our proof of Theorem~\ref{thm:SplayIsRepeatable},
Amit Halevi for comments that clarified the presentation of our execution
model, and Siddhartha Sen and Bernard Chazelle for editorial feedback. The
high-level presentation of Sections~\ref{sec:ApproximateMonotonicity}
and~\ref{sec:StartupOverhead} benefited from informal discussions with Daniel
Cooney. We are indebted to John Iacono for his guidance in understanding the
equivalence between Wilber's bound and Move-to-Root's crossing nodes, along
with corroborating our empirical comparisons between the behaviors of Splay and
Wilber's bound. Finally, we found David Galles' ``Data Structure
Visualizations'' website instrumental for prototyping our
proofs~\cite{VISUALIZER}. Research at Princeton University partially supported
by an innovation research grant from Princeton and a gift from Microsoft.
\end{acks}

\printbibliography
\end{document}